\documentclass{ifacconf}
\usepackage{amsmath}
\usepackage{amssymb}
\usepackage{graphicx}      
\usepackage{natbib}        
\usepackage{enumitem}
\usepackage{verbatim}
\usepackage{tikz}
\usepackage{cite}
\usepackage[caption=true]{subfig}
\newtheorem{ass}{Assumption}

\newtheorem{lemma}{Lemma}
\newtheorem{proposition}{Proposition}
\newtheorem{remark}{Remark}
\newtheorem{definition}{Definition}

\begin{document}
\begin{frontmatter}

\title{Multi-Agent Motion Planning and Object Transportation  under High Level Goals \thanksref{footnoteinfo}} 

\thanks[footnoteinfo]{This work was supported by the H2020 ERC Starting Grand BUCOPHSYS, the Swedish Research Council (VR), the Knut och Alice Wallenberg Foundation and the European Union's Horizon 2020 Research and Innovation Programme under the Grant Agreement No. 644128 (AEROWORKS).}

\author[Auth]{Christos K. Verginis} 
\author[Auth]{Dimos V. Dimarogonas}

\address[Auth]{KTH Royal Institute of Technology, 
   Stockholm, CO 10044 Sweden (e-mail: \{cverginis, dimos\}@kth.se)}

\begin{abstract}                
This paper presents a hybrid control framework for the motion planning of a multi-agent system including $N$ robotic agents and $M$ objects, under high level goals. In particular, we design control protocols that allow the transition of the agents as well as the transportation of the objects by the agents, among predefined regions of interest in the workspace. This allows us to abstract the coupled behavior of the agents and the objects as a finite transition system and to design a high-level multi-agent plan that satisfies the agents' and the objects' specifications, given as temporal logic formulas. Simulation results verify the proposed framework. 
\end{abstract}


\end{frontmatter}

\section{Introduction}
Temporal-logic based motion planning has gained significant amount of attention over the last decade, as it provides a fully automated  correct-by-design controller synthesis approach for autonomous robots. Temporal logics, such as linear temporal logic (LTL), provide formal high-level languages that can describe planning objectives more complex than the well-studied navigation algorithms, and have been used extensively both in single- as well as in multi-agent setups \citep{Fainekos2009,Lahijanian2016,Loizou2004,Diaz2015,Chen2012,Cowlagi2016,Belta2005,Bhatia2010,Bhatia2011,Filippidis2012,Meng15}.

Most works in the related literature consider temporal logic-based motion planning for fully actuated, autonomous agents. Consider, however, cases where some unactuated objects must undergo a series of processes in a workspace with autonomous agents (e.g., car factories). In such cases, the agents, except for satisfying their own motion specifications, are also responsible for coordinating with each other in order to transport the objects around the workspace. When the unactuated objects' specifications are expressed using temporal logics, then the abstraction of the agents' behavior becomes much more complex, since it has to take into account the objects' goals. 

Another issue regarding the temporal logic-based planning in the related literature is the non-realistic assumptions that are often considered. In particular, many works either do not take into account continuous agent dynamics or adopt single or double integrators \citep{Loizou2004,Filippidis2012,Fainekos2009,Bhatia2011,Meng15}, which can deviate from the actual dynamics of the agents, leading thus to poor performance in real-life scenarios. Moreover, many works adopt dimensionless point-mass agents and therefore do not consider inter-agent collision avoidance \citep{Belta2005,Filippidis2012,Meng15}, which can be a crucial safety issue in applications involving autonomous robots. 

This paper presents a novel hybrid control framework for the motion planning of a team of $N$ autonomous agents and $M$ unactuated objects under LTL specifications. We design feedback control laws for i) the navigation of the agents and ii) the transportation of the objects by the agents, among predefined regions of interest in the workspace, while ensuring inter-agent collision avoidance. This allows us to model the coupled behavior of the agents and the objects with a finite transition system, which can be used for the design of high-level plans that satisfy the given LTL specifications.  

\section{Notation and Preliminaries} \label{sec:Notation-and-Preliminaries}
Vectors and matrices are denoted with bold lowercase and uppercase letters, respectively, whereas scalars are denoted with non-bold lowercase letters.
The set of positive integers is denoted as $\mathbb{N}$ and the real $n$-space, with $n\in\mathbb{N}$, as $\mathbb{R}^n$;
$\mathbb{R}^n_{\geq 0}$ and $\mathbb{R}^n_{> 0}$ are the sets of real $n$-vectors with all elements nonnegative and positive, respectively, and $\mathbb{T}^n$ is the $n$-D torus. Given a set $S$, $2^S$ is the set of all possible subsets of $S$, $\lvert S \rvert$ is its cardinality, and, given a finite sequence $s_1,\dots,s_n$ of elements in $S$, with $n\in\mathbb{N}$, we denote by $(s_1,\dots,s_n)^\omega$ the infinite sequence $s_1,\dots,s_n,s_1,\dots,s_n,s_1,\dots$ created by repeating $s_1,\dots,s_n$. The notation $\|\boldsymbol{y}\|$ is used for the Euclidean norm of a vector $\boldsymbol{y} \in \mathbb{R}^n$. Given $x\in\mathbb{R}$ and $\boldsymbol{y},\boldsymbol{z}\in\mathbb{R}^n$, we use $\nabla_{\boldsymbol{z}}x = \partial x/\partial \boldsymbol{z}\in\mathbb{R}^n$ and  $\nabla_{\boldsymbol{z}}\boldsymbol{y} = \partial \boldsymbol{y}/\partial \boldsymbol{z}\in\mathbb{R}^{n\times n}$;
$\mathcal{B}_r(\boldsymbol{c})$ denotes the ball of radius $r\in\mathbb{R}_{>0}$ and center $\boldsymbol{c}\in\mathbb{R}^{3}$. Finally, we use $\mathcal{N}=\{1,\dots,N\},\mathcal{M}=\{1,\dots,M\},\mathcal{K}=\{1,\dots,K\}$, with $N,M,K\in\mathbb{N}$, as well as $\mathbb{M}=\mathbb{R}^3\times\mathbb{T}^3$. 

We focus on the task specification $\phi$ given as a Linear Temporal Logic (LTL) formula. The basic ingredients of a LTL formula are a set of atomic propositions $\mathcal{AP}$ and several boolean and temporal operators. LTL formulas are formed according to the following grammar \citep{baier2008principles}: $\phi ::= \mathsf{true}\: |\:a\: |\: \phi_{1} \land  \phi_{2}\: |\: \neg \phi\: |\:\bigcirc \phi\:|\:\phi_{1}\cup\phi_{2} $, where $a\in\mathcal{AP}$ and $\bigcirc$ (next), $\cup$ (until). Definitions of other useful operators like $\square$ (\it always\rm), $\lozenge$ (\it eventually\rm) and $\Rightarrow$ (\it implication\rm) are omitted and can be found at \citep{baier2008principles}.
The semantics of LTL are defined over infinite words over $2^{\mathcal{AP}}$. Intuitively, an atomic proposition $\psi\in\mathcal{AP}$ is satisfied on a word $w=w_1w_2\dots$ if it holds at its first position $w_1$, i.e. $\psi\in w_1$. Formula $\bigcirc\phi$ holds true if $\phi$ is satisfied on the word suffix that begins in the next position $w_2$, whereas $\phi_1\cup\phi_2$ states that $\phi_1$ has to be true until $\phi_2$ becomes true. Finally, $\lozenge\phi$ and  $\square\phi$ holds on $w$ eventually and always, respectively. For a full definition of the LTL semantics, the reader is referred to \citep{baier2008principles}.

\section{System Model and Problem Formulation} \label{sec:Model and PF}
Consider $N$ robotic agents operating in a workspace $\mathcal{W}$ with $M$ objects; $\mathcal{W}$ is a bounded sphere in $3$D space, i.e., $\mathcal{W}=\mathcal{B}_{r_0}(\boldsymbol{p}_0)=\{\boldsymbol{p}\in \mathbb{R}^3 \text{ s.t. } \lVert \boldsymbol{p}-\boldsymbol{p}_0 \rVert\leq r_0 \}$, where $\boldsymbol{p}_0\in \mathbb{R}^3$ and $r_0\in\mathbb{R}_{> 0}$ are the center and radius, respectively, of $\mathcal{W}$. The objects are represented by rigid bodies whereas the robotic agents are fully actuated and consist of a moving part (i.e., mobile base) and a robotic arm, having, therefore, access to the entire workspace.  Within $\mathcal{W}$ there exist $K$ smaller spheres around points of interest, which are described by $\mathcal{\pi}_k=\mathcal{B}_{r_k}(\boldsymbol{p}_{\pi_k})=\{\boldsymbol{p}\in \mathbb{R}^3 \text{ s.t. } \lVert \boldsymbol{p}-\boldsymbol{p}_{\pi_k} \rVert\leq r_k \}$, where $\boldsymbol{p}_{\pi_k}\in \mathbb{R}^3$ is the center and $r_k\in\mathbb{R}_{>0}$ the radius of $\pi_k$. The boundary of region $\pi_k$ is $\partial\pi_k = \left\{\boldsymbol{p}\in \mathbb{R}^3 \text{ s.t. }\lVert \boldsymbol{p}-\boldsymbol{p}_{\pi_k} \rVert = r_k\right\}$. We denote the set of all $\pi_k$ as $\Pi=\{\pi_1,\dots,\pi_K \}$. For the workspace partition to be valid, we consider that the regions of interest are sufficiently distant from each other and from the workspace boundary, i.e., $\lVert  \boldsymbol{p}_{\pi_k}-\boldsymbol{p}_{\pi_{k'}} \rVert > 4\max_{l\in\mathcal{K}}\{r_{\pi_l}\}$ and $\lVert \boldsymbol{p}_{\pi_k}-\boldsymbol{p}_0\rVert < r_0 - 3r_{\pi_k}, \forall k,k'\in\mathcal{K}$ with $k\neq k'$.      Moreover, we introduce disjoint sets of atomic propositions $\Psi_i, \Psi_{\scriptscriptstyle O_j}$, expressed as boolean variables, that represent services provided to agent $i\in\mathcal{N}$ and object $j\in\mathcal{M}$ in $\Pi$. The services provided at each region $\pi_k$ are given by the labeling functions $\mathcal{L}_i:\Pi\rightarrow2^{\Psi_i}, \mathcal{L}_{\scriptscriptstyle O_j}:\Pi\rightarrow2^{\Psi_{\scriptscriptstyle O_j}}$, which assign to each region $\pi_k, k\in\mathcal{K}$, the subset of services $\Psi_i$ and $\Psi_{\scriptscriptstyle O_j}$, respectively, that can be provided in that region to agent $i\in\mathcal{N}$ and object $j\in\mathcal{M}$, respectively.

\begin{figure}
\centering
\includegraphics[width = 0.25\textwidth]{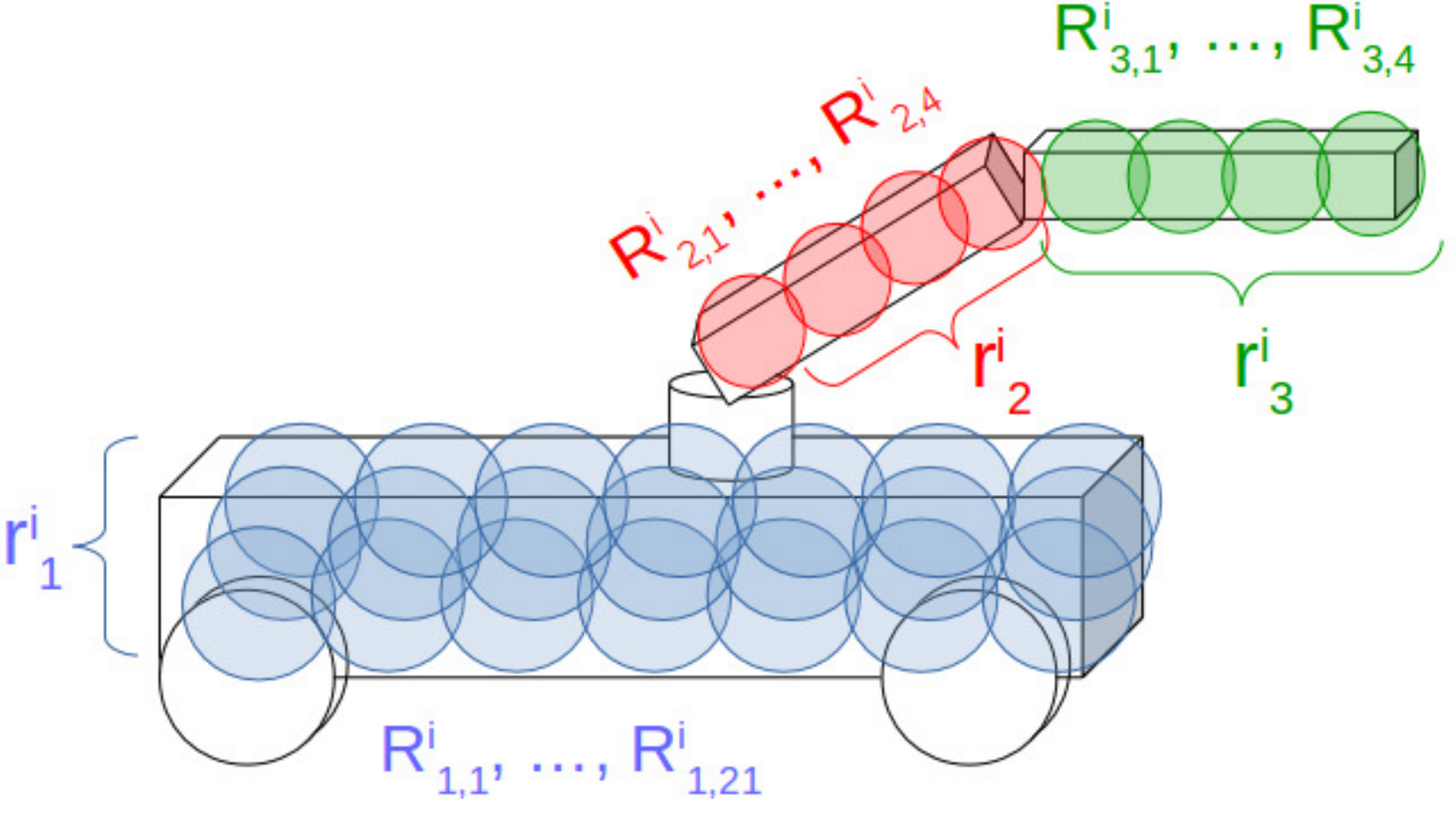}

\caption{ A robotic agent consisting of $\mathfrak{p}_i = 3$ rigid bodies, with $\mathfrak{R}^i_1 = 21$ and $\mathfrak{R}^i_2 = \mathfrak{R}^i_3 = 4$ ellipsoids. \label{fig:robot_rigid_ellips}}
\end{figure}

\subsection{System Model \label{subsec:system model}}

We denote by $\boldsymbol{q}_i:\mathbb{R}_{\geq 0}\rightarrow\mathbb{R}^{\mathfrak{n_i}},i\in\mathcal{N}$
the generalized joint variables of the $i$th agent, and $\boldsymbol{q} = [[\boldsymbol{q}^T_i]_{i\in\mathcal{N}}]^T\in\mathbb{R}^{\mathfrak{n}}$, with $\mathfrak{n} = \sum_{i\in\mathcal{N}}\mathfrak{n}_i$. We also denote as $\boldsymbol{p}_i:\mathbb{R}_{\geq 0}\rightarrow\mathbb{R}^3$ the position of the $i$th agent's end-effector derived from the forward kinematics \citep{Siciliano2010}, expressed in an inertial frame of reference.
The differential kinematics of agent $i$ suggest that $\boldsymbol{v}_i(t)  = \boldsymbol{J}_i(\boldsymbol{q}_i(t))\dot{\boldsymbol{q}}_i(t)$, where $\boldsymbol{J}_i:\mathbb{R}^{\mathfrak{n_i}}\rightarrow\mathbb{R}^{6\times\mathfrak{n_i}}$ is the Jacobian matrix and $\boldsymbol{v}_i:\mathbb{R}_{\geq 0}\to\mathbb{R}^6$, with $\boldsymbol{v}_i(t) = [\dot{\boldsymbol{p}}^T_i(t), \boldsymbol{\omega}^T_i(t)]^T$, is the velocity of the end-effector, with $\boldsymbol{\omega}_i:\mathbb{R}_{\geq 0}\to\mathbb{R}^{3}$ being its angular velocity with respect to (and expressed in) an inertial frame. The joint-space dynamics of agent $i$ are given by \citep{Siciliano2010}: 
\begin{equation}
\boldsymbol{B}_i(\boldsymbol{q}_i)\ddot{\boldsymbol{q}}_i + \boldsymbol{N}_i(\boldsymbol{q}_i, \dot{\boldsymbol{q}}_i)\dot{\boldsymbol{q}}_i + \boldsymbol{g}_i(\boldsymbol{q}_i) = \boldsymbol{\tau}_i - \boldsymbol{J}^T_i(\boldsymbol{q}_i) \boldsymbol{f}_i, \label{eq:joint space dynamics}
\end{equation}  
where $\boldsymbol{B}_i:\mathbb{R}^{\mathfrak{n_i}}\rightarrow\mathbb{R}^{\mathfrak{n_i} \times \mathfrak{n_i}}$ is the positive definite inertia matrix,  $\boldsymbol{N}_i:\mathbb{R}^{\mathfrak{n_i}} \times \mathbb{R}^{\mathfrak{n_i}} \rightarrow\mathbb{R}^{\mathfrak{n_i} \times \mathfrak{n_i}}$ is the Coriolis matrix, $\boldsymbol{g}_i:\mathbb{R}^{\mathfrak{n_i}}\rightarrow\mathbb{R}^{\mathfrak{n_{i}}}$ is the joint space gravity vector, $\boldsymbol{\tau}_i\in\mathbb{R}^{\mathfrak{n_{i}}}$ is the vector of joint torques and $\boldsymbol{f}_i\in\mathbb{R}^6$ is the vector of generalized forces that the end-effector exerts on a surface, in case of contact, $\forall i\in\mathcal{N}$.

We consider that each agent $i\in\mathcal{N}$ consists of $\mathfrak{p_i}$ rigid bodies $r_p^i,p\in \mathfrak{\bar{p}_i}$, where $\mathfrak{\bar{p}_i}=\{1,\cdots,\mathfrak{p_i}\}$ is the corresponding index set, whose volume $R_p^i$ is approximated by the union of $\mathfrak{R^i_p}$ generalized ellipsoids $R^i_{p,e}$, i.e., $R_p^i = \bigcup_{e\in\mathfrak{\bar{R}^i_p}} R^i_{p,e}$ with $\mathfrak{\bar{R}^i_p}=\{1,\cdots,\mathfrak{R^i_p}\}$. An example of the aforementioned formulation is illustrated in Fig. \ref{fig:robot_rigid_ellips}. 
By denoting the principal semi-axes lengths of $R^i_{p,e}$ as $a^i_{p,e},b^i_{p,e},c^i_{p,e}$,  
we define the function $\lambda^i_{p,e}: \mathbb{R}^3\times\mathbb{R}^{\mathfrak{n}_i}\rightarrow \mathbb{R}$ that describes $R^i_{p,e}$ as 
\begin{equation}
\lambda^i_{p,e}(\boldsymbol{p}^*,\boldsymbol{q}_i) = (\boldsymbol{p}^*)^T(\text{diag}\{a^i_{p,e},b^i_{p,e},c^i_{p,e}\})^{-1}\boldsymbol{p}^* - 1, \label{eq:ellipsoid}
\end{equation}
where $\boldsymbol{p}^* \in\mathbb{R}^3$ is a $3$D vector expressed in a local frame at the center of $R^i_{p,e}$, aligned with its principal axes, whose orientation depends on $\boldsymbol{q}_i$. We denote the solutions of \eqref{eq:ellipsoid} as $\Omega^{i,*}_{p,e}(\boldsymbol{q}_i) = \{\boldsymbol{p}\in\mathbb{R}^3 \text{ s.t. } \lambda^i_{p,e}(\boldsymbol{p},\boldsymbol{q}_i) = 0\}$. We also denote as $\Omega^{i,*}(\boldsymbol{q}_i) = \{\boldsymbol{p}\in\mathbb{R}^3 \text{ s.t. } (\exists e\in\mathfrak{\bar{R}^i_p},p\in \mathfrak{\bar{p}_i} \text{ s.t. } \lambda^i_{p,e}(\boldsymbol{p},\boldsymbol{q}_i) = 0) \}$, i.e., the set of solutions for all ellipsoids that approximate the volume of agent $i\in\mathcal{N}$.

Similarly to the robotic agents, we denote as $\boldsymbol{p}_{\scriptscriptstyle O_j}:\mathbb{R}_{\geq 0}\rightarrow \mathbb{R}^3, \boldsymbol{\eta}_{\scriptscriptstyle O_j}:\mathbb{R}_{\geq 0}\rightarrow\mathbb{T}^3$, the position and Euler-angle orientation of object $j\in\mathcal{M}$, which obeys the second order dynamics: 
\begin{equation}
\boldsymbol{M}_{\scriptscriptstyle O_j}(\boldsymbol{x}_{\scriptscriptstyle O_j})\dot{\boldsymbol{v}}_{\scriptscriptstyle O_j} + \boldsymbol{C}_{\scriptscriptstyle O_j}(\boldsymbol{x}_{\scriptscriptstyle O_j},\dot{\boldsymbol{x}}_{\scriptscriptstyle O_j})\boldsymbol{v}_{\scriptscriptstyle O_j} + \boldsymbol{g}_{\scriptscriptstyle O_j}(\boldsymbol{x}_{\scriptscriptstyle O_j}) = \boldsymbol{f}_{\scriptscriptstyle O_j}, \label{eq:object dynamics}
\end{equation}
where $\boldsymbol{x}_{\scriptscriptstyle O_j}=[\boldsymbol{p}^T_{\scriptscriptstyle O_j},\boldsymbol{\eta}^T_{\scriptscriptstyle O_j}]^T\in\mathbb{M}, \boldsymbol{v}_{\scriptscriptstyle O_j} = [\dot{\boldsymbol{p}}^T_{\scriptscriptstyle O_j},\boldsymbol{\omega}^T_{\scriptscriptstyle O_j}]^T\in\mathbb{R}^6, \boldsymbol{M}_{\scriptscriptstyle O_j}:\mathbb{M}\to\mathbb{R}^{6\times6}$ is the positive definite inertia matrix, $\boldsymbol{C}_{\scriptscriptstyle O_j}:\mathbb{M}\times\mathbb{R}^6\to\mathbb{R}^{6\times6}$ and $\boldsymbol{g}_{\scriptscriptstyle O_j}:\mathbb{M}\to\mathbb{R}^6$ are the Coriolis and gravity terms, respectively, and $\boldsymbol{f}_{\scriptscriptstyle O_j}\in \mathbb{R}^6$ is the vector of generalized forces acting on the object's center of mass. In case of a rigid grasp between agent $i$ and object $j$, $\boldsymbol{f}_i$ and $\boldsymbol{f}_{\scriptscriptstyle O_j}$ are related through $\boldsymbol{f}_i = \boldsymbol{G}_{i,j}(\boldsymbol{q}_i)\boldsymbol{f}_{\scriptscriptstyle O_j}$, where $\boldsymbol{G}_{i,j}:\mathbb{R}^{\mathfrak{}n_i}\rightarrow \mathbb{R}^{6\times6}$ is the full-rank grasp matrix. The aforementioned inertia and Coriolis matrices satisfy the skew-symmetric property \citep{Siciliano2010}:
\begin{subequations} \label{eq:skew_symm_property}
\begin{align}
\dot{\boldsymbol{B}}_i - 2\boldsymbol{N}_i =& - (\dot{\boldsymbol{B}}_i - 2\boldsymbol{N}_i)^T, \label{eq:skew_symm_property_agents} \\
\dot{\boldsymbol{M}}_{\scriptscriptstyle O_j} - 2\boldsymbol{C}_{\scriptscriptstyle O_j} =& - (\dot{\boldsymbol{M}}_{\scriptscriptstyle O_j} - 2\boldsymbol{C}_{\scriptscriptstyle O_j})^T, \label{eq:skew_symm_property_objects}
\end{align}
\end{subequations}
$\forall i\in\mathcal{N},j\in\mathcal{M}$.
 Object $j$ is a rigid body $r^j_{\scriptscriptstyle O}$ and therefore we approximate its volume $R^j_{\scriptscriptstyle O}$ by a union of $\mathfrak{R^j_{o} }$ ellipsoids, i.e., $R^j_{\scriptscriptstyle O} = \bigcup_{e\in\mathfrak{\bar{R}^j_{o} }} R^j_{\scriptscriptstyle O,e}$ with $\mathfrak{\bar{R}^j_{o} }=\{1,\cdots,\mathfrak{R^j_{o} }\}$ and $R^j_{\scriptscriptstyle O,e}$ 
 described by the functions $\lambda^j_{\scriptscriptstyle O,e}(\boldsymbol{x}_{\scriptscriptstyle O_j},\boldsymbol{p}^*)=(\boldsymbol{p}^*)^T(\text{diag}\{a^j_{\scriptscriptstyle O,e},b^j_{\scriptscriptstyle O,e},c^j_{\scriptscriptstyle O,e}\})^{-1}\boldsymbol{p}^*-1$, where 
 $a^j_{\scriptscriptstyle O,e},b^j_{\scriptscriptstyle O,e},c^j_{\scriptscriptstyle O,e}$ are the semi-axes lengths and $\boldsymbol{p}^*\in\mathbb{R}^3$ is a $3$D vector expressed in a local frame at the center of $R^j_{\scriptscriptstyle O,e}$, aligned with its principal axes.  Also, we denote $\Omega^{j,*}_{\scriptscriptstyle O,e}(\boldsymbol{x}_{\scriptscriptstyle O_j}) = \{\boldsymbol{p}\in\mathbb{R}^3 \text{ s.t. }\lambda^j_{\scriptscriptstyle O,e}(\boldsymbol{x}_{\scriptscriptstyle O_j},\boldsymbol{p}) = 0\}$ and $\Omega^{j,*}_{\scriptscriptstyle O}(\boldsymbol{x}_{\scriptscriptstyle O_j}) = \{\boldsymbol{p}\in\mathbb{R}^3 \text{ s.t. } (\exists e\in\mathfrak{\bar{R}^j_{o} } \text{ s.t. } \lambda^j_{\scriptscriptstyle O,e}(\boldsymbol{x}_{\scriptscriptstyle O_j},\boldsymbol{p}) = 0)\}$. 
 

We can now provide the following definitions:
\begin{definition} \label{def:agent in region}
Agent $i\in\mathcal{N}$ is in region $\pi_k,k\in\mathcal{K}$, at a configuration $\boldsymbol{q}_i$, denoted as $\mathcal{A}_i(\boldsymbol{q}_i)\in\pi_k$, iff $\lVert\boldsymbol{p} - \boldsymbol{p}_k\rVert \leq r_k - \varepsilon, \forall \boldsymbol{p}\in \Omega^{i,*}(\boldsymbol{q}_i)$, with $\varepsilon>0$ arbitrarily small.
\end{definition}

\begin{definition} \label{def:object in region}
Object $j\in\mathcal{M}$ is in region $\pi_k,k\in\mathcal{K}$, at a configuration $\boldsymbol{x}_{\scriptscriptstyle O_j}$, denoted as $\mathcal{O}_j(\boldsymbol{x}_{\scriptscriptstyle O_j})\in\pi_k$, iff $\lVert\boldsymbol{p} - \boldsymbol{p}_k\rVert \leq r_k - \varepsilon, \forall\boldsymbol{p} \in \Omega^{j,*}_{\scriptscriptstyle O}(\boldsymbol{x}_{\scriptscriptstyle O_j})$, with $\varepsilon>0$ arbitrarily small.
\end{definition}

In the following, we use the notation $\Omega^{i,*}(\boldsymbol{q}_i(t_0))\cap\Omega^{i',*}(\boldsymbol{q}_{i'}(t_0)) = \emptyset$,  $\Omega^{i,*}(\boldsymbol{q}_i(t_0))\cap\Omega^{j,*}_{\scriptscriptstyle O}(\boldsymbol{x}_{\scriptscriptstyle O_j}(t_0))= \emptyset,$ and $\Omega^{j,*}_{\scriptscriptstyle O}(\boldsymbol{x}_{\scriptscriptstyle O_j}(t_0))\cap\Omega^{j',*}_{\scriptscriptstyle O}(\boldsymbol{x}_{\scriptscriptstyle O_{j'}}(t_0))= \emptyset,
i,i'\in\mathcal{N},j,j'\in\mathcal{M}$, with $i\neq i', j\neq j'$, to describe collision-free cases at $t_0$ between the agents and the objects.

In order for our workspace discretization to be valid, we need the following assumption:
\begin{ass} \label{assumption}
There exist $\boldsymbol{q}^k_i, \boldsymbol{x}^k_{\scriptscriptstyle O_j}$ s. t. $\mathcal{A}(\boldsymbol{q}^k_i), \mathcal{O}(\boldsymbol{x}^k_{\scriptscriptstyle O_j})\in\pi_k$ and $\Omega^{i,*}(\boldsymbol{q}^k_i)\cap\Omega^{j,*}_{\scriptscriptstyle O}(\boldsymbol{x}^k_{\scriptscriptstyle O_j}) = \emptyset, \forall i\in\mathcal{N},j\in\mathcal{M},k\in\mathcal{K}$.   
\end{ass}
The aforementioned assumption implies that all regions of interest are sufficiently large to contain an object along with an agent, in a collision-free configuration. 

We also use the boolean variable $\mathcal{AG}_{i,j}(t^*)$ to denote whether agent $i\in\mathcal{N}$ rigidly grasps an object $j\in\mathcal{M}$ at the time instant $t^*$; $\mathcal{AG}_{i,0}(t^*)=\top$ denotes that agent $i$ does not grasp any object. Note that $\mathcal{AG}_{i,\ell}(t^*)=\top, \ell\in\{0\}\cup\mathcal{M} \Leftrightarrow \mathcal{AG}_{i,j}(t^*)=\bot, \forall j\in\{0\}\cup\mathcal{M}\backslash\{\ell\}$ (i.e., agent $i$ can grasp at most one object at a time) and $\Omega^{i,*}(\boldsymbol{q}_i(t^*))\cap\Omega^{j,*}_{\scriptscriptstyle O}(\boldsymbol{x}_{\scriptscriptstyle O_j}(t^*))=\emptyset, \forall j\in\mathcal{M},\Rightarrow\mathcal{AG}_{i,0}(t)=\top$. The following definitions address the transitions of the agents and the objects between the regions of interest. 

\begin{definition}  \label{def:agent transition}
Assume for agent $i\in\mathcal{N}$ that $\mathcal{A}_i(\boldsymbol{q}_i(t_0))\in\pi_k, k\in\mathcal{K}$, and
\begin{enumerate}[label=(\roman*)]
\item $\Omega^{i,*}(\boldsymbol{q_i}(t_0))\cap\left(\Omega^{j,*}_{\scriptscriptstyle O}(\boldsymbol{x}_{\scriptscriptstyle O_j}(t_0))\cup\Omega^{n,*}(\boldsymbol{q}_n(t_0))\right)=\emptyset$, 
\end{enumerate}
$\forall j\in\mathcal{M},n\in\mathcal{N}\backslash\{i\}$, for some $t_0\in\mathbb{R}_{\geq 0}$. Then, there exists a transition for agent $i$ from region $\pi_k$ to $\pi_{k'},k'\in\mathcal{K}$, denoted as $\pi_k\rightarrow_i\pi_{k'}$, iff there exists a finite $t_f\in\mathbb{R}_{\geq 0}$ with $t_f \geq t_0$ and a bounded control trajectory $\boldsymbol{\tau}_i:[t_0,t_f]\rightarrow\mathbb{R}^{\mathfrak{n}_i}$ such that $\mathcal{A}_i(\boldsymbol{q}_i(t_f))\in\pi_{k'}$ and
\begin{equation}
\Omega^{i,*}(\boldsymbol{q}_i(t))\cap\left(\partial\pi_m\cup\Omega^{n,*}(\boldsymbol{q}_n(t))\cup\Omega^{j,*}_{\scriptscriptstyle O}(\boldsymbol{x}_{\scriptscriptstyle O_j}(t) )\right)= \emptyset, \notag
\end{equation}
$\forall t\in[t_0, t_f], j\in\mathcal{M}, n\in\mathcal{N}\backslash\{i\},m\in\mathcal{K}\backslash\{k,k'\}$.
\end{definition}

\begin{definition} \label{def:grasping }
Assume for agent $i\in\mathcal{N}$ and object $j\in\mathcal{M}$ that $\mathcal{A}_i(\boldsymbol{q}_i(t_0)), \mathcal{O}_j(\boldsymbol{x}_{\scriptscriptstyle O_j}(t_0))\in\pi_k,k\in\mathcal{K}$, and 
\begin{enumerate}[label=(\roman*)]
\item $\Omega^{i,*}(\boldsymbol{q}_i(t_0))\cap\left(\Omega^{j',*}_{\scriptscriptstyle O}(\boldsymbol{x}_{\scriptscriptstyle O_{j'}}(t_0))\cup \Omega^{n,*}(\boldsymbol{q}_n(t_0)) \right)=\emptyset$,
\item $\Omega^{j,*}_{\scriptscriptstyle O}(\boldsymbol{x}_{\scriptscriptstyle O_j}(t_0))\cap\left(\Omega^{\ell,*}_{\scriptscriptstyle O}(\boldsymbol{x}_{\scriptscriptstyle O_\ell}(t_0))\cup \Omega^{i',*}(\boldsymbol{q}_{i'}(t_0))  \right) =\emptyset$,
\end{enumerate}
$\forall \ell\in\mathcal{M}\backslash\{j\},n\in\mathcal{N}\backslash\{i\},j'\in\mathcal{M},i'\in\mathcal{N}$, for some $t_0\in\mathbb{R}_{\geq 0}$. Then, agent $i$ \textit{grasps} object $j$ at region $\pi_k$, denoted as $i\xrightarrow{g}j$, iff there exists a finite $t_f\in\mathbb{R}_{\geq 0}$ with $t_f \geq t_0$ and a bounded control trajectory $\boldsymbol{\tau}_i:[t_0,t_f]\rightarrow\mathbb{R}^{\mathfrak{n}_i}$ such that $\mathcal{AG}_{i,j}(t_f)=\top, \mathcal{A}_i(\boldsymbol{q}_i(t)), \mathcal{O}_j(\boldsymbol{x}_{\scriptscriptstyle O_j}(t))\in\pi_k$ and 
\begin{enumerate}[label=(\roman*)]
\item $\Omega^{i,*}(\boldsymbol{q}_i(t))\cap \left(\Omega^{n,*}(\boldsymbol{q}_n(t)) \cup \Omega^{\ell,*}_{\scriptscriptstyle O}(\boldsymbol{x}_{\scriptscriptstyle O_\ell}(t)) \right)=\emptyset$,
\item $\Omega^{j,*}_{\scriptscriptstyle O}(\boldsymbol{x}_{\scriptscriptstyle O_j}(t))\cap \left(\Omega^{\ell,*}_{\scriptscriptstyle O}(\boldsymbol{x}_{\scriptscriptstyle O_\ell}(t))\cup \Omega^{n,*}(\boldsymbol{q}_n(t))\right)=\emptyset$,
\end{enumerate}
 $\forall t\in[t_0,t_f], n\in\mathcal{N}\backslash\{i\},\ell\in\mathcal{M}\backslash\{j\}$.
\end{definition}
The action of an agent \textit{releasing} a rigid grasp with an object at a region, denoted as $i\xrightarrow{r}j$, is defined similarly and is omitted.
\begin{definition} \label{def:agent-object transition}
Assume for agent $i\in\mathcal{N}$ and object $j\in\mathcal{M}$ that $\mathcal{A}_i(\boldsymbol{q_i}(t_0)), \mathcal{O}_j(\boldsymbol{x}_{\scriptscriptstyle O_j}(t_0))\in\pi_k$,  $k\in\mathcal{K}$, with $\mathcal{AG}_{i,j}(t_0)=\top$ and
\begin{enumerate}[label=(\roman*)]
\item  $\Omega^{i,*}(\boldsymbol{q}_i(t_0))\cap\left(\Omega^{n,*}(\boldsymbol{q}_n(t_0))\cup \Omega^{\ell,*}_{\scriptscriptstyle O}(\boldsymbol{x}_{\scriptscriptstyle O_\ell}(t_0)) \right)=\emptyset$.
\item  $\Omega^{j,*}_{\scriptscriptstyle O}(\boldsymbol{x}_{\scriptscriptstyle O_j}(t_0))\cap\left(\Omega^{\ell,*}_{\scriptscriptstyle O}(\boldsymbol{x}_{\scriptscriptstyle O_\ell}(t_0))\cup \Omega^{n,*}(\boldsymbol{q}_n(t_0)) \right)=\emptyset$,
\end{enumerate}
$\forall n\in\mathcal{N}\backslash\{i\},\ell\in\mathcal{M}\backslash\{j\}$  for some $t_0\in\mathbb{R}_{\geq 0}$. Then, agent $i$ transports object $j$ from region $\pi_k$ to region $\pi_{k'}, k'\in\mathcal{K}$, denoted as $\pi_k \xrightarrow{T}_{i,j} \pi_{k'}$, if there exists a finite $t_f\in\mathbb{R}_{\geq 0}$ with $t_f \geq t_0$ and a bounded control trajectory $\boldsymbol{\tau}_i:[t_0,t_f]\rightarrow\mathbb{R}^{\mathfrak{n}_i}$ such that $\mathcal{A}_i(\boldsymbol{q}_i(t_f)), \mathcal{O}_j(\boldsymbol{x}_{\scriptscriptstyle O_j}(t_f))\in\pi_{k'}$, $\mathcal{AG}_{i,j}(t) = \top$, and
\begin{enumerate}[label=(\roman*)]
\item $\left(\Omega^{i,*}(\boldsymbol{q}_i(t))\cup\Omega^{j,*}_{\scriptscriptstyle O}(\boldsymbol{x}_{\scriptscriptstyle O_j}(t))\right)  \cap\partial\pi_m = \emptyset$,
\item $\Omega^{i,*}(\boldsymbol{q}_i(t))\cap \left(\Omega^{n,*}(\boldsymbol{q}_n(t))\cup \Omega^{\ell,*}_{\scriptscriptstyle O}(\boldsymbol{x}_{\scriptscriptstyle O_\ell}(t))\right) = \emptyset$,
\item $\Omega^{j,*}_{\scriptscriptstyle O}(\boldsymbol{x}_{\scriptscriptstyle O_j}(t))\cap\left(\Omega^{\ell,*}_{\scriptscriptstyle O}(\boldsymbol{x}_{\scriptscriptstyle O_\ell}(t))\cup \Omega^{n,*}(\boldsymbol{q}_n(t))\right)= \emptyset$, 
\end{enumerate} 
$\forall t\in[t_0,t_f], n\in\mathcal{N}\backslash\{i\},\ell\in\mathcal{M}\backslash\{j\}, m\in\mathcal{K}\backslash\{k,k'\}$.
\end{definition}


\subsection{Specification} \label{subsec:Specification}
Our goal is to control the multi-agent system such that the agents and the objects obey a given specification over their atomic propositions $\Psi_i, \Psi_{\scriptscriptstyle O_j}, \forall i\in\mathcal{N},j\in\mathcal{M}$. 
Given the trajectories $\boldsymbol{q}_i(t), \boldsymbol{x}_{\scriptscriptstyle O_j}(t), t\in\mathbb{R}_{\geq 0}$, of agent $i$ and object $j$, respectively, their corresponding \textit{behaviors} are given by the infinite sequences $\beta_i = (\boldsymbol{q}_i(t),\sigma_i)=(\boldsymbol{q}_i(t_{i_1}),\sigma_{i_1})(\boldsymbol{q}_i(t_{i_2}),\sigma_{i_2})\dots, \beta_{\scriptscriptstyle O_j} = (\boldsymbol{x}_{\scriptscriptstyle O_j}(t),\sigma_{\scriptscriptstyle O_j})=\\(\boldsymbol{x}_{\scriptscriptstyle O_j}(t_{\scriptscriptstyle O_{j,1}}),\sigma_{\scriptscriptstyle O_{j,1}}) (\boldsymbol{x}_{\scriptscriptstyle O_j}(t_{\scriptscriptstyle O_{j,2}}),\sigma_{\scriptscriptstyle O_{j,2}})\dots$ with $t_{i_{m+1}} > t_{i_{m}} \geq 0, t_{\scriptscriptstyle O_{j,m+1}} > t_{\scriptscriptstyle O_{j,m}} \geq 0, \forall m\in\mathbb{N}$. The sequences $\sigma_i, \sigma_{\scriptscriptstyle O_j}$ are the services provided to the agent and the object, respectively, over their trajectories, i.e., $\sigma_{i_m}\in 2^{\Psi_i}, \sigma_{\scriptscriptstyle O_{j,l}}\in 2^{\Psi_{\scriptscriptstyle O_j}}$ with $\mathcal{A}_i(\boldsymbol{q}_i(t_{i_m}))\in \pi_{k_m}, \sigma_{i_m} \in \mathcal{L}_i(\pi_{k_m})$ and $\mathcal{O}_j(\boldsymbol{x}_{\scriptscriptstyle O_j}(t_{\scriptscriptstyle O_{j,l}}))\in \pi_{k_l}, \sigma_{\scriptscriptstyle O_{j,l}} \in \mathcal{L}_{\scriptscriptstyle O_j}(\pi_{k_l}), k_m, k_l \in\mathcal{K}, \forall m,l\in\mathbb{N}$, with $\mathcal{L}_i$ and $\mathcal{L}_{\scriptscriptstyle O_j}$ as defined in Section \ref{sec:Model and PF}.
 
\begin{definition}
The behaviors $\beta_i, \beta_{\scriptscriptstyle O_j}$ satisfy formulas $\phi_i, \phi_{\scriptscriptstyle O_j}$ iff $\sigma_i \models \phi_i$ and $\sigma_{\scriptscriptstyle O_j} \models \phi_{\scriptscriptstyle O_j}$, respectively.
\end{definition}

\subsection{Problem Formulation}
The control objectives are given as LTL formulas $\phi_i, \phi_{\scriptscriptstyle O_j}$ over $\Psi_i, \Psi_{\scriptscriptstyle O_j}$, respectively, $\forall i\in\mathcal{N},j\in\mathcal{M}$. The LTL formulas $\phi_i, \phi_{\scriptscriptstyle O_j}$ are satisfied if there exist behaviors $\beta_i,\beta_{\scriptscriptstyle O_j}$ of agent $i$ and object $j$ that satisfy  $\phi_i, \phi_{\scriptscriptstyle O_j}$. Formally, the problem treated in this paper is the following:
\begin{prob} \label{problem}
Consider $N$ robotic agents and $M$ objects in $\mathcal{W}$ subject to the dynamics (\ref{eq:joint space dynamics}) and (\ref{eq:object dynamics}), respectively, and (i) $\Omega^{i,*}(\boldsymbol{q}_i(0))\cap\Omega^{j,*}_{\scriptscriptstyle O}(\boldsymbol{x}_{\scriptscriptstyle O_j}(0))=\emptyset$,  (ii) $\dot{\boldsymbol{q}}_i(0) = \boldsymbol{0}, \forall i\in\mathcal{N}$, (iii) $\mathcal{A}_i(\boldsymbol{q}_i(0))\in\pi_{i_0}, \mathcal{O}_j(\boldsymbol{x}_{\scriptscriptstyle O_j}(0) )\in\pi_{\scriptscriptstyle O_{j,0}}$, with $i_0, O_{j,0}\in\mathcal{K}$ and $i_0\neq {n}_0, O_{j,0}\neq O_{\ell,0}, \forall i,n\in\mathcal{N},j,\ell\in\mathcal{M}$ with $i\neq n, j\neq \ell$.  Given the disjoint set $\Psi_i,\Psi_{\scriptscriptstyle O_j}$, $N$ LTL formulas $\phi_i$ over $\Psi_i$ and $M$ LTL formulas $\phi_{\scriptscriptstyle O_j}$ over $\Psi_{\scriptscriptstyle O_j}$, develop a control strategy that achieves behaviors $\beta_i, \beta_{\scriptscriptstyle O_j}$ which yield the satisfaction of 
$\phi_i, \phi_{\scriptscriptstyle O_j}, \forall i\in\mathcal{N},j\in\mathcal{M}$.
\end{prob}

\section{Main Results}

\subsection{Continuous Control Design}
The first ingredient of our solution is the development of feedback control laws that establish agent transitions and object transportations as defined in Def. \ref{def:agent transition} and \ref{def:agent-object transition}, respectively. Regarding the grasping actions of Def. \ref{def:grasping }, we employ one of the already existing methodologies that can derive the corresponding control laws (e.g., \citep{Cutkosky2012},\citep{Reis2015}). 

\subsubsection{\bf Transformation to Point World: \rm}

In this work we employ the algorithm proposed in \citep{Tanner2003} to create point worlds. In particular, there exists a sequence of smooth transformations on the rigid body ellipsoids, introduced in Section \ref{sec:Model and PF}, that creates spaces where the robotic agents and the objects are represented by points. Since $3$D spheres are a special case of $3$D ellipsoids, we also consider the regions of interest as obstacles that will be transformed to points. For details on the transformation, the reader is referred to \citep{Tanner2003}. 

Assume that the conditions of Problem \ref{problem} hold for some $t_0\in\mathbb{R}_{\geq 0}$, i.e., all agents and objects are located in regions of interest and there is no more than one agent or one object at the same region.
We design a control law such that a subset of agents performs a transition between two regions of interest and another subset of agents performs object transportation, according to Def. \ref{def:agent transition} and \ref{def:agent-object transition}, respectively. 
Let $\mathcal{Z}, \mathcal{V}, \mathcal{G}, \mathcal{Q}\subseteq \mathcal{N}$ denote disjoint sets of agents corresponding to transition, transportation, grasping and releasing actions, respectively, with $\lvert \mathcal{V} \rvert + \lvert \mathcal{G} \rvert + \lvert \mathcal{Q} \rvert  \leq \lvert \mathcal{M} \rvert$ and $\mathcal{A}_z(\boldsymbol{q}_z(t_0))\in\pi_{k_z},\mathcal{A}_v(\boldsymbol{q}_v(t_0))\in\pi_{k_v}, \mathcal{A}_g(\boldsymbol{q}_g(t_0))\in\pi_{k_g}, \mathcal{A}_q(\boldsymbol{q}_q(t_0))\in\pi_{k_q}, k_z,k_v,k_g,k_q\in\mathcal{K}, \forall z\in\mathcal{Z}, v\in\mathcal{V},g\in\mathcal{G},q\in\mathcal{Q}$. Note that there might be idle agents in some regions, not performing any actions, i.e., $\mathcal{Z}\cup\mathcal{V}\cup\mathcal{G}\cup\mathcal{Q} \subseteq \mathcal{N}$. Let also $\mathcal{S}=\{[s_v]_{v\in\mathcal{V}}\}, \mathcal{X}=\{[x_g]_{g\in\mathcal{G}}\}, \mathcal{Y}=\{[y_q]_{q\in\mathcal{Q}}\}\subseteq\mathcal{M}$ such that $\mathcal{O}_{s_v}(\boldsymbol{x}_{\scriptscriptstyle O_{s_v}}(t_0))\in\pi_{k_v}, \mathcal{O}_{x_g}(\boldsymbol{x}_{\scriptscriptstyle O_{x_g}}(t_0))\in\pi_{k_g}, \mathcal{O}_{y_q}(\boldsymbol{x}_{\scriptscriptstyle O_{y_q}}(t_0))\in\pi_{k_y}, s_v\in\mathcal{S}, x_g\in\mathcal{X}, y_g\in\mathcal{Y}, \forall v\in\mathcal{V}, g\in\mathcal{G}, q\in\mathcal{Q}$, i.e., there exists one object at each $\pi_{k_v}, \pi_{k_g}, \pi_{k_q}$. Moreover, assume that the conditions of Def. \ref{def:agent transition} hold for all $z\in\mathcal{Z}$, the conditions of Def. \ref{def:agent-object transition} hold for all $v\in\mathcal{V}$ and  $s_v\in\mathcal{S}$, the conditions of Def. \ref{def:grasping } hold for all $g\in\mathcal{G}$ and $x_g\in\mathcal{X}$, and the corresponding \text{release} conditions (which are omitted due to space limitations), hold for all $q\in\mathcal{Q}$ and $y_q\in\mathcal{Y}$.    
In the following, we design $\boldsymbol{\tau}_{z}$ and $\boldsymbol{\tau}_{v}$ such that $\pi_{k_z}\rightarrow_z\pi_{k'_z}$ and $\pi_{k_v}\xrightarrow{T}_{v,s_v}\pi_{k'_v},k'_z,k'_v\in\mathcal{K}, \forall z\in\mathcal{Z},v\in\mathcal{V}$, assuming that there exist appropriate $\boldsymbol{\tau}_g$ and $\boldsymbol{\tau}_q$ that guarantee $g\xrightarrow{g}x_g$ and $q\xrightarrow{r}y_q$ in $\pi_g, \pi_q$, respectively. 

Regarding the transitions $\pi_{k_z}\rightarrow\pi_{k'_z}$ by agents $z\in\mathcal{Z}$, we define the error function $\gamma_{z,{k'_z}}:\mathbb{R}^{\mathfrak{n}_z}\rightarrow \mathbb{R}_{\geq 0}$ with $\gamma_{z,{k'_z}}(\boldsymbol{q}_z) = \lVert \boldsymbol{p}_z(\boldsymbol{q}_z) - \boldsymbol{p}_{\pi_{k'_z}} \rVert^2$. 

Regarding the transportations $\pi_{k_v}\xrightarrow{T}_{v,s_v}\pi_{k'_v}$ of the objects $s_v\in\mathcal{S}$ by agents $v\in\mathcal{V}$, note first that a rigid grasp between agent $v$ and object $s_v$ creates a continuous dependence of $\boldsymbol{x}_{\scriptscriptstyle O_{s_v}}$ 
on $\boldsymbol{q}_v$. Therefore, we can write $\boldsymbol{p}_{\scriptscriptstyle O_{s_v}}(t) = \boldsymbol{k}_{v,s_v}(\boldsymbol{q}_v(t))$, where $\boldsymbol{k}_{v,s_v}:\mathbb{R}^{\mathfrak{n}_v}\rightarrow\mathbb{R}^3$ can be considered as the forward kinematics to the object's center of mass. 
Therefore, we define the error function $\gamma^{s_v}_{v,{k'_v}}:\mathbb{R}^{\mathfrak{n}_v}\rightarrow \mathbb{R}_{\geq 0}$ as $\gamma^{s_v}_{v,{k'_v}}(\boldsymbol{q_v}) =  \lVert \boldsymbol{k}_{v,s_v}(\boldsymbol{q}_v) - \boldsymbol{p}_{\pi_{k'_v}}\rVert^2$.  

Each agent should avoid (i) collision with other agents and objects and (ii) entering other regions of interest except from its goal one, both in the transition and transportation actions. Consider the agents $z\in\mathcal{Z}$ and $v\in\mathcal{V}$ performing $\pi_{k_z}\rightarrow_z\pi_{k'_z}$ and $\pi_{k_v}\xrightarrow{T}_{v,s_v}\pi_{k'_v}$, respectively. For each rigid body $r^z_p$ of $z$, all other rigid bodies $r_{p'}^z, p'\in\mathfrak{\bar{p}}_z\backslash\{p\}, r_{p}^{z'}, p\in\mathfrak{\bar{p}}_{z'}, z'\in\mathcal{V}\cup\mathcal{Z}\backslash\{z\}$ and objects $r^j_{\scriptscriptstyle O},\forall j\in\mathcal{M}\backslash\{\mathcal{X}\cup\mathcal{Y}\}$ as well as the spheres $\mathcal{B}_{r_m}(\boldsymbol{p}_{\pi_m}),m\in\mathcal{K}\backslash\{k_z,k'_z\}$, are considered as obstacles for $r_p^z$. In the same vein, for each rigid body $r^v_p$ of $v$, all other rigid bodies $r_{p'}^v, p'\in\mathfrak{\bar{p}}_v\backslash\{p\}, r_{p}^{v'}, p\in\mathfrak{\bar{p}}_{v'}, v'\in\mathcal{V}\cup\mathcal{Z}\backslash\{v\}$ as well as the spheres $\mathcal{B}_{r_m}(\boldsymbol{p}_{\pi_m}),m\in\mathcal{K}\backslash\{k_v,k'_v\}$, are considered as obstacles for $r_p^v$. Note that collision avoidance with the set of agents $\mathcal{G},\mathcal{Q}$ that perform grasp and release actions does not need to be considered, since Def. \ref{def:grasping } implies that these agents are contained in $\pi_{k_g},\pi_{k_q}$ during their action and therefore, the avoidance of $\pi_{k_g},\pi_{k_q}$ is sufficient. 

\textit{Singularity Avoidance:} Singularity regions are sets of measure zero within the joint space that depend on the mechanical structure of the agent. The singularity space for agent $i\in\mathcal{N}$  is defined as $Q^s_{i} = \{\boldsymbol{q}_i\in\mathbb{R}^{\mathfrak{n_i}} \text{ s.t. } \det(\boldsymbol{J}^T_i(\boldsymbol{q}_i)\boldsymbol{J}_i(\boldsymbol{q}_i))=0 \}$. In well-designed manipulators, singularities can be decoupled to classes that depend on a subset of the joint variables. Therefore, we can enclose these regions inside ellipsoids representing artificial obstacles that affect the motion of the robot end-effector. 

Hence, each rigid body $r^i_p$ of agent $i\in\{z,v\},z\in\mathcal{Z},v\in\mathcal{V}$ has a number $N^i_{p,\text{obs}}$ obstacles to avoid, including the other rigid bodies, the undesired regions of interest and the singularity ellipsoids, as analyzed above. Then, by  employing the point world transformation algorithm of \citep{Tanner2003}, $r^i_p$ is transformed to the point $\boldsymbol{h}^i_p(\boldsymbol{q}_i)\in\mathbb{R}^3$ and the obstacles are transformed to the points $\boldsymbol{h}^{i,\text{obs}}_{p,o}(\widetilde{\boldsymbol{q}})\in\mathbb{R}^3,\forall o\in\bar{N}^i_{p,\text{obs}}= \{1,\dots,N^i_{p,\text{obs}}\}$, where $\widetilde{\boldsymbol{q}} = [[\boldsymbol{q}^T_z]_{z\in\mathcal{Z}}, [\boldsymbol{q}^T_v]_{v\in\mathcal{V}}]^T\in\mathbb{R}^{\widetilde{\mathfrak{n}}},  \widetilde{\mathfrak{n}} = \sum_{z\in\mathcal{Z}}\mathfrak{n}_z+\sum_{v\in\mathcal{V}}\mathfrak{n}_v$. 

To form the ``obstacle" function, we adopt the notion of proximity relations of \citep{Loizou2006}, which are all the possible collision schemes between the aforementioned transformed points. A measure of the distance for each $r^i_p$ and its obstacles is the function $\beta^{i,\text{obs}}_{p,o}:\mathbb{R}^{\widetilde{\mathfrak{n}}}\rightarrow \mathbb{R}_{\geq 0}$ with $\beta^{i,\text{obs}}_{p,o}(\widetilde{\boldsymbol{q}})= \lVert \boldsymbol{h}^i_p(\boldsymbol{q}_i) - \boldsymbol{h}^{i,\text{obs}}_{p,o}(\widetilde{\boldsymbol{q}})\rVert^2, o\in\bar{N}^i_{p,\text{obs}}$. By considering the relation proximity function, which represents the sum of all distance measures in a specific relation  between the transformed points, we can define the relation verification function (RVF), as in \citep{Loizou2006}. Then, the total ``obstacle" function $\beta_{\text{obs}}:\mathbb{R}^{\widetilde{\mathfrak{n}}}\rightarrow \mathbb{R}_{\geq 0}$ is the product of the RVFs for all relations and resembles the possible collision schemes between all $r^i_p, i\in\{z,v\},z\in\mathcal{Z},v\in\mathcal{V}$, and the corresponding obstacles. For more details on the technique, the reader is referred to \citep{Loizou2006}.
Regarding the workspace boundaries, we form the function $\delta^i_{p,e}:\mathbb{R}^{\mathfrak{n_i}}\rightarrow \mathbb{R}_{\geq 0}$ with $\delta^i_{p,e}(\boldsymbol{q}_i)= (r_0 - \max\{a^i_{p,e}, b^i_{p,e}, c^i_{p,e}\})^2-\lVert \boldsymbol{p}_{\scriptscriptstyle R^i_{p,e}}(\boldsymbol{q}_i) - \boldsymbol{p}_0\rVert^2$, that represents an over-approximation of the distance of ellipsoid $R^i_{p,e}$ from the workspace boundary, with $\boldsymbol{p}_{\scriptscriptstyle R^i_{p,e}}$ being the ellipsoid's center. Then, $\delta^i:\mathbb{R}^{\mathfrak{n}_i}\rightarrow \mathbb{R}_{\geq 0}$, with $\delta^i(\boldsymbol{q}_i)=\prod_{p\in\bar{\mathfrak{p}}_i,e\in\bar{\mathfrak{R}}^i_p }  \delta^i_{p,e}(\boldsymbol{q}_i)$, encodes the distance of agent $i$ from the workspace boundaries.

We construct now the following multi-agent navigation function $\varphi:\mathbb{R}^{\widetilde{\mathfrak{n}}}\rightarrow [0,1]$ \citep{Koditchek92,Loizou2006}, that incorporates the desired behavior of the agents:
\begin{equation}
\varphi(\widetilde{\boldsymbol{q}}) = \frac{\gamma(\widetilde{\boldsymbol{q}})} { ( \gamma^\kappa(\widetilde{\boldsymbol{q}}) + \beta_{\text{obs}}(\widetilde{\boldsymbol{q}})\prod_{i\in\mathcal{Z}\cup\mathcal{V}}\delta^i(\boldsymbol{q}_i) \ )^\frac{1}{\kappa}}, \label{eq:navigation_function}
\end{equation}
where $\kappa \in\mathbb{R}_{>0}$ and $\gamma: \mathbb{R}^{\widetilde{\mathfrak{n}}}\rightarrow \mathbb{R}_{\geq 0}$ is defined as 
$\small \gamma(\widetilde{\boldsymbol{q}}) = \sum_{z\in\mathcal{Z}}\gamma_{z,k'_z}(\boldsymbol{q_z}) + \sum_{v\in\mathcal{V}}\gamma^{s_v}_{v,k'_v}(\boldsymbol{q}_v)$. Note that, a sufficient condition for avoidance of the undesired regions and avoidance of collisions and singularities is $\varphi < 1$.

Next, we design the control protocols $\boldsymbol{\tau}_z:[t_0,\infty)\rightarrow\mathbb{R}^{\mathfrak{n}_z}, \boldsymbol{\tau}_v:[t_0,\infty)\rightarrow\mathbb{R}^{\mathfrak{n}_v}$: 
\begin{subequations} \label{eq:control_protocol}
\begin{align}
\boldsymbol{\tau}_z(t) =& \boldsymbol{g}_z(\boldsymbol{q}_z) - \nabla_{\boldsymbol{q}_z}\varphi(\widetilde{\boldsymbol{q}}) - \boldsymbol{K}_z\dot{\boldsymbol{q}}_z(t), 	\label{eq:control_protocol_transition} \\
\boldsymbol{\tau}_v(t) =& \bar{\boldsymbol{g}}_{v,s_v}(\boldsymbol{q}_v) - \nabla_{\boldsymbol{q}_v}\varphi(\widetilde{\boldsymbol{q}}) - \boldsymbol{K}_v\dot{\boldsymbol{q}}_v(t),  	\label{eq:control_protocol_transportation}
\end{align}
\end{subequations}
$\forall z\in\mathcal{Z}, v\in\mathcal{V}$, where $\boldsymbol{K}_i=\text{diag}\{k_i\}\in\mathbb{R}^{\mathfrak{n_i}\times\mathfrak{n_i}}$, with $k_i\in\mathbb{R}_{> 0}$, is a constant positive definite gain matrix, $\forall i\in\{z,v\},z\in\mathcal{Z},v\in\mathcal{V}$, and 
\begin{equation}
\bar{\boldsymbol{g}}_{i,j} = \boldsymbol{g}_i + \bar{\boldsymbol{G}}^T_{i,j}\boldsymbol{g}_{\scriptscriptstyle O_j}, \notag
\end{equation}
is the coupled agent-object gravity vector $\forall i\in\mathcal{N},j\in\mathcal{M}$. In the same vein, we also define the coupled matrices  
\begin{align*}
\bar{\boldsymbol{B}}_{i,j} =& \ \boldsymbol{B}_i + \bar{\boldsymbol{G}}^T_{i,j} \boldsymbol{M}_{\scriptscriptstyle O_{j}} \bar{\boldsymbol{G}}_{i,j}, \\ 
\bar{\boldsymbol{N}}_{i,j} =& \ \boldsymbol{N}_i + \bar{\boldsymbol{G}}^T_{i,j} \boldsymbol{M}_{\scriptscriptstyle O_{j}} \dot{\bar{\boldsymbol{G}}}_{i,j} + (\bar{\boldsymbol{G}}_{i,j})^T \boldsymbol{C}_{\scriptscriptstyle O_j} \bar{\boldsymbol{G}}_{i,j}, \\ 
\bar{\boldsymbol{G}}_{i,j} =& \ \boldsymbol{G}^T_{i,j}\boldsymbol{J}_i, 
\end{align*}
$\forall i\in\mathcal{N},j\in\mathcal{M}$. The following Proposition is needed for the subsequent analysis:
\begin{proposition} \label{prop:skew_symm}
The matrix $\bar{\boldsymbol{B}}_{i,j}$ is positive definite and the matrix $\dot{\bar{\boldsymbol{B}}}_{i,j} - 2\bar{\boldsymbol{N}}_{i,j}$ is skew-symmetric, $\forall i\in\mathcal{N},j\in\mathcal{M}$.
\end{proposition}
Proof: The positive definiteness of $\bar{\boldsymbol{B}}_{i,j}$ can be deduced from the positive definiteness of $\boldsymbol{B}_i$ and $\boldsymbol{M}_{\scriptscriptstyle O_j}$.  
Moreover, by defining $\boldsymbol{A} = \dot{\bar{\boldsymbol{G}}}^T_{i,j} \boldsymbol{M}_{\scriptscriptstyle O_j}\bar{\boldsymbol{G}}_{i,j}$, we obtain:
\begin{equation}
\small
\dot{\bar{\boldsymbol{B}}}_{i,j} - 2\bar{\boldsymbol{N}}_{i,j} = \dot{\boldsymbol{B}}_i - 2\boldsymbol{N}_i+ \bar{\boldsymbol{G}}^T_{i,j}(\boldsymbol{M}_{\scriptscriptstyle O_j} - 2\boldsymbol{C}_{\scriptscriptstyle O_j} ) + \boldsymbol{A} - \boldsymbol{A}^T, \notag
\normalsize
\end{equation} 
from which, by employing \eqref{eq:skew_symm_property}, we obtain $(\dot{\bar{\boldsymbol{B}}}_{i,j} - 2\bar{\boldsymbol{N}}_{i,j})^T = -(\dot{\bar{\boldsymbol{B}}}_{i,j} - 2\bar{\boldsymbol{N}}_{i,j})$, which completes the proof.

\begin{lemma} \label{lem:agent transition}
Consider the sets of agent $\mathcal{Z}, \mathcal{V}$ and the set of objects $\mathcal{S}$ as defined above, described by the dynamics \eqref{eq:joint space dynamics} and \eqref{eq:object dynamics}, such that the conditions of Def. \ref{def:agent transition} hold for all $z\in\mathcal{Z}$ and the conditions of Def. \ref{def:agent-object transition} hold for all $v\in\mathcal{V}, s_v\in\mathcal{S}$ for $t_0=0, k_z, k_v\in\mathcal{K}$. Then the control protocols \eqref{eq:control_protocol} guarantee that $\pi_{k_z}\rightarrow_z\pi_{k'_z}$ and $\pi_{k_v}\xrightarrow{T}_{v,s_v}\pi_{k'_v},k'_z,k'_v\in\mathcal{K}, \forall z\in\mathcal{Z}, v\in\mathcal{V}$, according to Def. \ref{def:agent transition} and \ref{def:agent-object transition}, respectively.
\end{lemma}
\begin{pf}
Regarding the agents $v\in\mathcal{V}$, note that, since $\boldsymbol{f}_v = \boldsymbol{G}_{v,s_v}(\boldsymbol{q}_v)\boldsymbol{f}_{\scriptscriptstyle O_{s_v}}$ owing to the rigid grasp between agent $v$ and object $s_v$, the kineto-statics duality \citep{Siciliano2010} suggests that 
\begin{equation}
\boldsymbol{v}_{\scriptscriptstyle O_{s_v}}(t) = \boldsymbol{G}^T_{v,s_v}(\boldsymbol{q}_v) \boldsymbol{v}_v(t) = \boldsymbol{G}^T_{v,s_v}(\boldsymbol{q}_v)\boldsymbol{J}_v(\boldsymbol{q}_v)\dot{\boldsymbol{q}}_v. \label{eq:rigid duality}
\end{equation} 
By substituting \eqref{eq:object dynamics} in $\boldsymbol{f}_v = \boldsymbol{G}_{v,s_v}(\boldsymbol{q}_v)\boldsymbol{f}_{\scriptscriptstyle O_{s_v}}$ and using the derivative of \eqref{eq:rigid duality}, we obtain
\begin{align}
\boldsymbol{J}^T_v \boldsymbol{f}_v =& \bar{\boldsymbol{G}}^T_{v,s_v}\boldsymbol{M}_{\scriptscriptstyle O_{s_v}}\bar{\boldsymbol{G}}_{v,s_v}\ddot{\boldsymbol{q}}_v + \left(\bar{\boldsymbol{G}}^T_{v,s_v}\boldsymbol{M}_{\scriptscriptstyle O_{s_v}}\dot{\bar{\boldsymbol{G}}}_{v,s_v} + \notag \right.\\
&\left. \bar{\boldsymbol{G}}^T_{v,s_v}\boldsymbol{C}_{\scriptscriptstyle O_{s_v}}\bar{\boldsymbol{G}}_{v,s_v} \right)\dot{\boldsymbol{q}}_v + \bar{\boldsymbol{G}}^T_{v,s_v}\boldsymbol{g}_{\scriptscriptstyle O_{s_v}}. \label{eq:coupled_proof}
\end{align}
Then, by substituting \eqref{eq:coupled_proof} in \eqref{eq:joint space dynamics}, we obtain:
\begin{equation}
\bar{\boldsymbol{B}}_{v,s_v}(\boldsymbol{q}_v)\ddot{\boldsymbol{q}}_v + \bar{\boldsymbol{N}}_{v,s_v}(\boldsymbol{q}_v,\dot{\boldsymbol{q}}_v)\dot{\boldsymbol{q}}_v + \bar{\boldsymbol{g}}_{v,s_v}(\boldsymbol{q}_v) = \boldsymbol{\tau}_v, \label{eq:coupled_dynamics}
\end{equation}
i.e., the coupled dynamics of agent $v$ and object $s_v$.

Next, we consider the Lyapunov function $V:\mathbb{R}^{\widetilde{\mathfrak{n}}}\times\mathbb{R}^{\widetilde{\mathfrak{n}}}\rightarrow \mathbb{R}_{\geq 0}$, with
$\small V(\widetilde{\boldsymbol{q}},\dot{\widetilde{\boldsymbol{q}}}) = \varphi(\boldsymbol{\tilde{\boldsymbol{q}}}) + \sum_{z\in\mathcal{Z}}\tfrac{1}{2}\dot{\boldsymbol{q}}^T_zB_z(\boldsymbol{q}_z)\dot{\boldsymbol{q}}_z +  \sum_{v\in\mathcal{V}}\tfrac{1}{2}\dot{\boldsymbol{q}}^T_v\bar{B}_{v,s_v}(\boldsymbol{q}_v)\dot{\boldsymbol{q}}_v \normalsize$. Note that, since no collisions occur at $t=0$ and $\dot{\boldsymbol{q}}_i(0) = \boldsymbol{0}, \forall i\in\mathcal{N}$, we conclude from \eqref{eq:navigation_function} that $V_0 = V(\widetilde{\boldsymbol{q}}(0),\dot{\widetilde{\boldsymbol{q}}}(0)) < 1$. By differentiating $V$ with respect to time,
and substituting \eqref{eq:joint space dynamics} and \eqref{eq:coupled_dynamics}, we obtain:	
\begin{align}
\dot{V} = \sum\limits_{z\in\mathcal{Z}}&\{ (\nabla_{\boldsymbol{q}_z} \varphi)^T\dot{\boldsymbol{q}}_z + \dot{\boldsymbol{q}}^T_z(\boldsymbol{\tau}_z - \boldsymbol{N}_z\dot{\boldsymbol{q}}_z - \boldsymbol{g}_z) + \tfrac{1}{2}\dot{\boldsymbol{q}}^T_z\dot{\boldsymbol{B}}_z \dot{\boldsymbol{q}}_z\} \notag \\
&+ \sum\limits_{v\in\mathcal{V}}\{ (\nabla_{\boldsymbol{q}_v} \varphi)^T\dot{\boldsymbol{q}}_v + \dot{\boldsymbol{q}}^T_v(\boldsymbol{\tau}_v - \bar{\boldsymbol{N}}_{v,s_v}\dot{\boldsymbol{q}}_v - \bar{\boldsymbol{g}}_{v,s_v}) + \notag\\ & \tfrac{1}{2}\dot{\boldsymbol{q}}^T_v\dot{\bar{\boldsymbol{B}}}_{v,s_v} \dot{\boldsymbol{q}}_v \}, \notag
\end{align}
where we have also used the fact that $\boldsymbol{f}_z = 0,\forall z\in\mathcal{Z}$, since the agents performing transportation actions are not in contact with any objects or other agents.  
By employing \eqref{eq:skew_symm_property}, \eqref{eq:control_protocol}, as well as Proposition \ref{prop:skew_symm}, $\dot{V}$ becomes:
\begin{align}
\dot{V} =& -\sum\limits_{i\in\mathcal{Z}\cup\mathcal{V}} \dot{\boldsymbol{q}}^T_i \boldsymbol{K}_i \dot{\boldsymbol{q}}_i, \notag
\end{align}
which is strictly negative unless $\dot{\boldsymbol{q}}_i = \boldsymbol{0}, \forall i\in\{z,v\},z\in\mathcal{Z},v\in\mathcal{V}$. Hence, $V(\widetilde{\boldsymbol{q}}(t),\dot{\widetilde{\boldsymbol{q}}}(t)) \leq V_0 < 1 \Rightarrow \varphi(\widetilde{\boldsymbol{q}}(t)) < 1, \forall t\in\mathbb{R}_{\geq 0}$, which implies that collisions, undesired  regions and singularities are avoided.  Moreover, according to La Salle's Invariance Principle \citep{Khalil}, the system will converge to the largest invariant set contained in $\{\widetilde{\boldsymbol{q}},\dot{\widetilde{\boldsymbol{q}}}, \text{ s.t. } \dot{\widetilde{\boldsymbol{q}}} = \boldsymbol{0}\}$. In order for this set to be invariant, we require that $\ddot{\widetilde{\boldsymbol{q}}} = \boldsymbol{0}$, which, by employing \eqref{eq:control_protocol} and the dynamics \eqref{eq:joint space dynamics}, \eqref{eq:coupled_dynamics}, implies that $\nabla_{\boldsymbol{q}_i}\varphi = \boldsymbol{0}, \forall i\in\{z,v\},z\in\mathcal{Z},v\in\mathcal{V}$. Since $\varphi$ is a navigation function \citep{Koditchek92}, the condition $\nabla_{\boldsymbol{q}_i}\varphi = \boldsymbol{0}$ is true only at the destination configuration and a set of isolated saddle points. By choosing $\kappa$ sufficiently large, the region of attraction of the saddle points is a set of measure zero \citep{Koditchek92}. Thus, the system converges to the destination configuration from almost everywhere, i.e., $\lVert \boldsymbol{p}_z(\boldsymbol{q}_z) - \boldsymbol{p}_{\pi_{k'_z}}\rVert \rightarrow 0$ and $\lVert \boldsymbol{k}_{v,s_v}(\boldsymbol{q}_v) - \boldsymbol{p}_{\pi_{k'_v}}\rVert \rightarrow 0$. Therefore, in view of Assumption \ref{assumption}, there exist finite time instants $t_{f_z},t_{f_v} > t_0, \forall z\in\mathcal{Z},v\in\mathcal{V}$, such that $\mathcal{A}_z(\boldsymbol{q}_z(t_{f_z}))\in\pi_{k'_z}$ and $\mathcal{A}_v(\boldsymbol{q}_v(t_{f_v})),\mathcal{O}_{s_v}(\boldsymbol{x}_{\scriptscriptstyle O_{s_v}}(t_{f_v}))\in\pi_{k'_v}$, with inter-agent collision and singularity avoidance. Thus, $\pi_{k_z}\rightarrow_z\pi_{k'_z}$ and $\pi_{k_v}\xrightarrow{T}_{v,s_v}\pi_{k'_v}$ according to Def. \ref{def:agent transition} and \ref{def:agent-object transition}, respectively, which completes the proof. 
\end{pf}

\begin{remark}
During the transitions $\pi_{z_k}\rightarrow_z\pi_{k'_z}, z\in\mathcal{Z}$, once agent $z$ leaves $\pi_{k_z}$, there is no guarantee that it will not enter it again until it reaches $\pi_{k'_z}$. The same holds for $\pi_v\xrightarrow{T}_{v,s_v}\pi_{v'}, v\in\mathcal{V}$, as well. For that reason, we can modify \eqref{eq:navigation_function} to include continuous switchings to a navigation controllers that avoid $\pi_{k_z}$ (or $\pi_{k_v}$), once agent $z$ (or $v$) is out ot it, as in \citep{Meng15}.
\end{remark}

Considering the agents $g\in\mathcal{G},q\in\mathcal{Q}$ that perform grasp and release actions, note that there exist positive and finite time instants $t_{f_g}, t_{f_q} > t_0$ that these actions will be completed, $\forall g\in\mathcal{G},q\in\mathcal{Q}$. We define $\bar{t}_f = \max\{  \max_{g\in\mathcal{G}} \{t_{f_g}\},\max_{q\in\mathcal{Q}} \{t_{f_q}\}, \max\mathbb{S}\}$, where $\mathbb{S} = \{t \geq t_0 \text{ s.t. } \mathcal{A}_z(\boldsymbol{q}_z(t))\in\pi_{k'_z}, \mathcal{A}_v(\boldsymbol{q}_v(t)), \mathcal{O}_{s_v}(\boldsymbol{x}_{\scriptscriptstyle O_{s_v}}(t))\in\pi_{k'_v}, \forall z\in\mathcal{Z}, v\in\mathcal{V} \}$, which represents a time instant that all the agents $i\in\mathcal{N}$ will have completed their respective action. Therefore, by choosing $t'_0 > \bar{t}_f$, we can define a new set of actions to be executed by the agents, starting at $t'_0$ (i.e., the conditions of Def. \ref{def:agent transition}-\ref{def:agent-object transition} hold at $t'_0$ instead of $t_0$). In this way, we add a notion of synchronization to our system, since each (non-idle) agent, after completing an action, will wait for all other agents to complete their own, so that they start the next set of actions at the same time.   

\subsection{High-Level Plan Generation} \label{sec:high level plan}

The second part of the solution is the derivation of a high-level plan that satisfies the given LTL formulas $\phi_i$ and $\phi_{\scriptscriptstyle O_j}$ and can be generated by using standard techniques from automata-based formal verification methodologies. Thanks to (i) the proposed control laws that allow agent transitions and object transportations $\pi_k\rightarrow_i\pi_{k'}$ and $\pi_k\xrightarrow{T}_{i,j}\pi_{k'}$, respectively, (ii) the off-the-self control laws that guarantee grasp and release actions $i\xrightarrow{g}j$ and $i\xrightarrow{r}j$, respectively, and (iii) the formulation for the synchronization of actions, we can abstract the behavior of the agents using a finite transition system as presented in the sequel.

\begin{definition} \label{def:TS objects all agents}
The coupled behavior of the overall system of all the $N$ agents and $M$ objects is modeled by the transition system $\mathcal{TS} = (\Pi_s,\Pi^{\text{init}}_s, \rightarrow_{s},\mathcal{AG}, \Psi, \mathcal{L} )$,
where \\
\textbf{(i)} $\Pi_s\subset \bar{\Pi}\times\bar{\Pi}_{\scriptscriptstyle O}\times\bar{\mathcal{AG}}$ is the set of states; 
$\bar{\Pi}=\Pi_1\times\cdots\times\Pi_N$ and $\bar{\Pi}_{\scriptscriptstyle O}=\Pi_{\scriptscriptstyle O_1}\times\cdots\times\Pi_{\scriptscriptstyle O_M}$ are the set of states-regions that the agents and the objects can be at, with $\Pi_i  = \Pi_{\scriptscriptstyle O_j} = \Pi,\forall i\in\mathcal{N},j\in\mathcal{M}$; 
$\mathcal{AG}= \mathcal{AG}_1\times\cdots\times\mathcal{AG}_N$ is the set of boolean grasping variables introduced in Section \ref{subsec:system model}, with
$\mathcal{AG}_i = \{\mathcal{AG}_{i,0}\}\cup\{[\mathcal{AG}_{i,j}]_{j\in\mathcal{M}}\}, \forall i\in\mathcal{N}$. 
By denoting $\bar{\pi}=\left(\pi_{k_1},\cdots,\pi_{k_N}\right),\bar{\pi}_{\scriptscriptstyle O}=(\pi_{\scriptscriptstyle k_{\scriptscriptstyle O_1}},\cdots,\pi_{\scriptscriptstyle k_{\scriptscriptstyle O_M}}), \bar{w}=\left(w_1,\cdots,w_N\right)$, with $\pi_{k_i},\pi_{k_{\scriptscriptstyle O_j}}\in\Pi$ (i.e., $k_i,k_{\scriptscriptstyle O_j}\in\mathcal{K},\forall i\in\mathcal{N},j\in\mathcal{M}$) and $w_i\in\mathcal{AG}_i, \forall i\in\mathcal{N}$, then $\pi_s = (\bar{\pi},\bar{\pi}_{\scriptscriptstyle O},\bar{w})\in\Pi_s$ iff $\pi_{k_i}\neq\pi_{k_n}$ and $\pi_{k_{\scriptscriptstyle O_j}}\neq\pi_{k_{\scriptscriptstyle O_{\ell}}}, \forall i,n\in\mathcal{N},j,\ell\in\mathcal{M}$, with $i\neq n$ and $j \neq \ell$, i.e., we consider that there cannot be more than one agent or more than one object at a time in each region of interest, \\
\textbf{(ii)} $\Pi^{\text{init}}_s\subset\Pi_s$ is the initial set of states at $t=0$, which, owing to \textbf{(i)}, satisfies the conditions of Problem \ref{problem},\\
\textbf{(iii)} $\rightarrow_s\subset \Pi_s\times\Pi_s$ is a transition relation defined as follows: given the states $\pi_s, \pi'_s\in\Pi$, with
\small
\begin{align}
\pi_s =& (\bar{\pi},\bar{\pi}_{\scriptscriptstyle O},\bar{w}) = (\pi_{k_1},\dots,\pi_{k_N}, \pi_{k_{\scriptscriptstyle O_1}},\dots,\pi_{k_{\scriptscriptstyle O_M}},w_1,\dots,w_N), \notag \\
\pi'_s =& ( \bar{\pi}',\bar{\pi}'_{\scriptscriptstyle O},\bar{w}') = (\pi_{k'_1},\dots,\pi_{k'_N}, \pi_{k'_{\scriptscriptstyle O_1}},\dots,\pi_{k'_{\scriptscriptstyle O_1}},w'_1,\dots,w'_N), \label{eq:pi_s}
\end{align}
\normalsize
a transition $\pi_s \rightarrow_s \pi'_s $ occurs iff there exist disjoint sets $\mathcal{Z},\mathcal{V},\mathcal{G},\mathcal{Q}\subseteq \mathcal{N}$ with $\lvert \mathcal{V} \rvert + \lvert \mathcal{G} \rvert + \lvert \mathcal{Q} \rvert  \leq \lvert M \rvert $ and $\mathcal{S} = \{[s_v]_{v\in\mathcal{V}}\}, \mathcal{X} = \{[x_g]_{g\in\mathcal{G}}\}, \mathcal{Y} = \{[y_q]_{q\in\mathcal{Q}}\}\subseteq\mathcal{M}$, such that:
\begin{enumerate}
\item $w_z = w'_z = \mathcal{AG}_{z,0} = \top$ and $\pi_{k_z}\rightarrow_z \pi_{k'_z}, \forall z\in\mathcal{Z}$,
\item $\pi_{k_v} = \pi_{k_{\scriptscriptstyle O_{s_v}}}, \pi_{k'_v} = \pi_{k'_{\scriptscriptstyle O_{s_v}}}, w_v = w'_v = \mathcal{AG}_{v,s_v} =\top  $  and $\pi_{k_v}\xrightarrow{T}_{v,s_v} \pi_{k'_v}, s_v\in\mathcal{S}, \forall v\in\mathcal{V}$.
\item $\pi_{k_g} = \pi_{k'_g} = \pi_{k_{\scriptscriptstyle O_{x_g}}} = \pi_{k'_{\scriptscriptstyle O_{x_g}}},   w_g = \mathcal{AG}_{g,0} = \top, w'_g = \mathcal{AG}_{g,x_g} = \top$ and $g\xrightarrow{g}x_g, x_g\in\mathcal{X},\forall g\in\mathcal{G}$,
\item $\pi_{k_{q}} = \pi_{k'_{q}} = \pi_{k_{\scriptscriptstyle O_{y_{q}}}} = \pi_{k'_{\scriptscriptstyle O_{y_{q}}}},   w_{q} = \mathcal{AG}_{q,y_q} = \top, w'_{q} = \mathcal{AG}_{q,0} = \top$ and $q\xrightarrow{r}y_{q}, y_{q}\in\mathcal{Y},\forall q\in\mathcal{Q}$, \\
\end{enumerate}
\textbf{(iv)} $\Psi = \bar{\Psi}\cup\bar{\Psi}_{\scriptscriptstyle O}$ with $\bar{\Psi}=\bigcup_{i\in\mathcal{N}}\Psi_{i}$ and $\Psi_{\scriptscriptstyle O} = \bigcup_{j\in\mathcal{M}}\Psi_{\scriptscriptstyle O_j}$, are the atomic propositions of the agents and objects, respectively, as defined in Section \ref{sec:Model and PF}, \\
\textbf{(v)} $\mathcal{L}:\Pi_s \rightarrow 2^\Psi$ is a labeling function defined as follows: Given a state $\pi_s$ as in \eqref{eq:pi_s} and $\psi = \bigcup_{i\in\mathcal{N}}\psi_i\bigcup_{j\in\mathcal{M}}\psi_{\scriptscriptstyle O_j}$ with $\psi_i\in2^{\Psi_i},\psi_{\scriptscriptstyle O_j}\in2^{\Psi_{\scriptscriptstyle O_j}}$, then $\psi\in\mathcal{L}(\pi_s)$ if and only if $\psi_i\in\mathcal{L}_i(\pi_{k_i})$ and $\psi_{\scriptscriptstyle O_j}\in\mathcal{L}_{\scriptscriptstyle O_j}(\pi_{k_{\scriptscriptstyle O_j}}), \forall i\in\mathcal{N},j\in\mathcal{M}$.
\end{definition}

Broadly speaking, the cases (1)-(4) in the transition relation of Def. \ref{def:TS objects all agents} correspond to agent transition, object transportation, grasp and release actions, respectively.  More specifically, the agents $z\in\mathcal{Z}$ perform transitions between $\pi_{k_z}, \pi_{k'_z}$, the agents $v\in\mathcal{V}$ perform transportation of the objects $s_v\in\mathcal{S}$ from $\pi_{k_v}$ to $\pi_{k'_v}$, and the agents $g\in\mathcal{G}, q\in\mathcal{Q}$ perform grasping and releasing actions with objects $x_g\in\mathcal{X}$ and $y_q\in\mathcal{Y}$, respectively. 

Next, we form the global LTL formula $\phi = (\land_{i\in\mathcal{N}}\phi_i)\land(\land_{j\in\mathcal{M}}\phi_{\scriptscriptstyle O_j})$ over the set $\Psi$. Then, we translate $\phi$ to a Buchi Automaton $\mathcal{BA}$ and 
we build the product $\widetilde{\mathcal{TS}} = \mathcal{TS}\times\mathcal{BA}$. The accepting runs of $\widetilde{\mathcal{TS}}$ satisfy $\phi$ and are directly projected to a sequence of desired states to be visited in $\mathcal{TS}$. Although the semantics of LTL are defined over infinite sequences of services, it can be proven that there always exists a high-level plan that takes the form of a finite state sequence followed by an infinite repetition of another finite state sequence. For more details on the followed technique, the reader is referred to the related literature, e.g., \citep{baier2008principles}.

Following the aforementioned methodology, we obtain a high-level plan as sequences of states and atomic propositions $p = \pi^1_s\pi^2_s\dots$ and $\psi = \psi^1 \psi^2\dots$, with $\pi^m_s = (\bar{\pi}^m,\bar{\pi}^m_{\scriptscriptstyle O},\bar{w}^m)\in\Pi_s, \psi^m  \in 2^{\Psi}, \psi^m\in\mathcal{L}(\pi^m_s)$, $\forall m\in\mathbb{N}$, and $\psi \models \phi$. 
The path $p$ is then projected to individual sequences of regions $\pi_{k^1_{\scriptscriptstyle O_j}} \pi_{k^2_{\scriptscriptstyle O_j}}\dots$ with $\pi_{k^m_{\scriptscriptstyle O_j}}\in\Pi,\forall m\in\mathbb{N}$,  $\pi_{k^1_i} \pi_{k^2_i}\dots$ with $\pi_{k^m_i}\in\Pi,\forall m\in\mathbb{N}$, and boolean grasping variables $w^1_iw^2_i\dots$ with $w^m_i\in\mathcal{AG}_{i}, \forall m\in\mathbb{N},i\in\mathcal{N},j\in\mathcal{M}$. The aforementioned sequences determine the behavior of agent $i\in\mathcal{N}$, i.e., the sequence of actions (transition, transportation, grasp, release or stay idle) it must take.

By the definition of $\mathcal{L}$ in Def. \ref{def:TS objects all agents}, we obtain that $\psi^m_i\in\mathcal{L}_i(\pi_{k^m_i}), \psi^m_{\scriptscriptstyle O_j}\in\mathcal{L}_{\scriptscriptstyle O_j}(\pi_{k^m_{\scriptscriptstyle O_j}}), \forall i\in\mathcal{N},j\in\mathcal{M},m\in\mathbb{N}$. Therefore, since $\phi = (\land_{i\in\mathcal{N}}\phi_i)\land(\land_{j\in\mathcal{M}}\phi_{\scriptscriptstyle O_j})$ is satisfied by $\psi$, we conclude that $\psi_i=(\psi^1_i\psi^2_i\dots) \models \phi_i$ and $\psi_{\scriptscriptstyle O_j}=(\psi^1_{\scriptscriptstyle O_j}\psi^2_{\scriptscriptstyle O_j}\dots) \models \phi_{\scriptscriptstyle O_j}, \forall i\in\mathcal{N}, j\in\mathcal{M}$.


The sequence of the states $(\pi_{k^1_i}\pi_{k^2_i}\dots,$ $\psi^1_i\psi^2_i\dots)$ and $(\pi_{k^1_{\scriptscriptstyle O_j}}\pi_{k^2_{\scriptscriptstyle O_j}}\dots, \psi^1_{\scriptscriptstyle O_j}\psi^2_{\scriptscriptstyle O_j}\dots)$ over $(\Pi, 2^{\Psi_i})$ and $(\Pi, 2^{\Psi_{\scriptscriptstyle O_j}})$, respectively, produces the trajectories $\boldsymbol{q}_i(t)$ and $\boldsymbol{x}_{\scriptscriptstyle O_j}(t), \forall i\in\mathcal{N},j\in\mathcal{M}$. The corresponding behaviors are $\beta_i = (\boldsymbol{q}_i(t),\sigma_i) = (\boldsymbol{q}_i(t_{i_1}),\sigma_{i_1})(\boldsymbol{q}_i(t_{i_2}),\sigma_{i_2})\dots$ and $\beta_{\scriptscriptstyle O_j} = (\boldsymbol{x}_{\scriptscriptstyle O_j}(t),\sigma_{\scriptscriptstyle O_j})= (\boldsymbol{x}_{\scriptscriptstyle O_j}(t_{\scriptscriptstyle O_{j,1}}),\sigma_{\scriptscriptstyle O_{j,1}})(\boldsymbol{x}_{\scriptscriptstyle O_j}(t_{\scriptscriptstyle O_{j,2}}),\sigma_{\scriptscriptstyle O_{j,2}})\dots$, respectively, according to Section \ref{subsec:Specification}, with $\mathcal{A}_i(\boldsymbol{q}_i(t_{i_m}))\in\pi_{k^m_i}, \sigma_{i_m}\in\mathcal{L}_i(\pi_{k^m_i})$ and $\mathcal{O}_j(\boldsymbol{x}_{\scriptscriptstyle O_j}(t_{\scriptscriptstyle O_{j,m}}))\in\pi_{k^m_{\scriptscriptstyle O_j}}, \sigma_{\scriptscriptstyle O_{j,m}}\in\mathcal{L}_{\scriptscriptstyle O_j}(\pi_{k^m_{\scriptscriptstyle O_j}})$. Thus, it is guaranteed that $\sigma_i \models \phi_i,\sigma_{\scriptscriptstyle O_j} \models \phi_{\scriptscriptstyle O_j}$ and consequently, the behaviors $\beta_i$ and $\beta_{\scriptscriptstyle O_j}$ satisfy the formulas $\phi_i$ and $\phi_{\scriptscriptstyle O_j}$, respectively, $\forall i\in\mathcal{N},j\in\mathcal{M}$. The aforementioned reasoning is summarized as follows:
\begin{thm}
The execution of the path $(p,\psi)$ of $\mathcal{TS}$ guarantees behaviors $\beta_i,\beta_{\scriptscriptstyle O_j}$ that yield the satisfaction of  $\phi_i$ and $\phi_{\scriptscriptstyle O_j}$, respectively, $\forall i\in\mathcal{N},j\in\mathcal{M}$, providing, therefore, a solution to Problem \ref{problem}.  
\end{thm}

\begin{remark}
Note that although the overall set of states of $\mathcal{TS}$ increases exponentially with respect to the number of agents/objects/regions (the maximum number of states is $K^{N+M}(M+1)^N$), some states are either not reachable or simply removed due to the constraints set of Def. \ref{def:TS objects all agents}, reducing the state complexity.
\end{remark}

\begin{figure}
\centering
\includegraphics[width = 0.4\textwidth, height = 0.2\textheight]{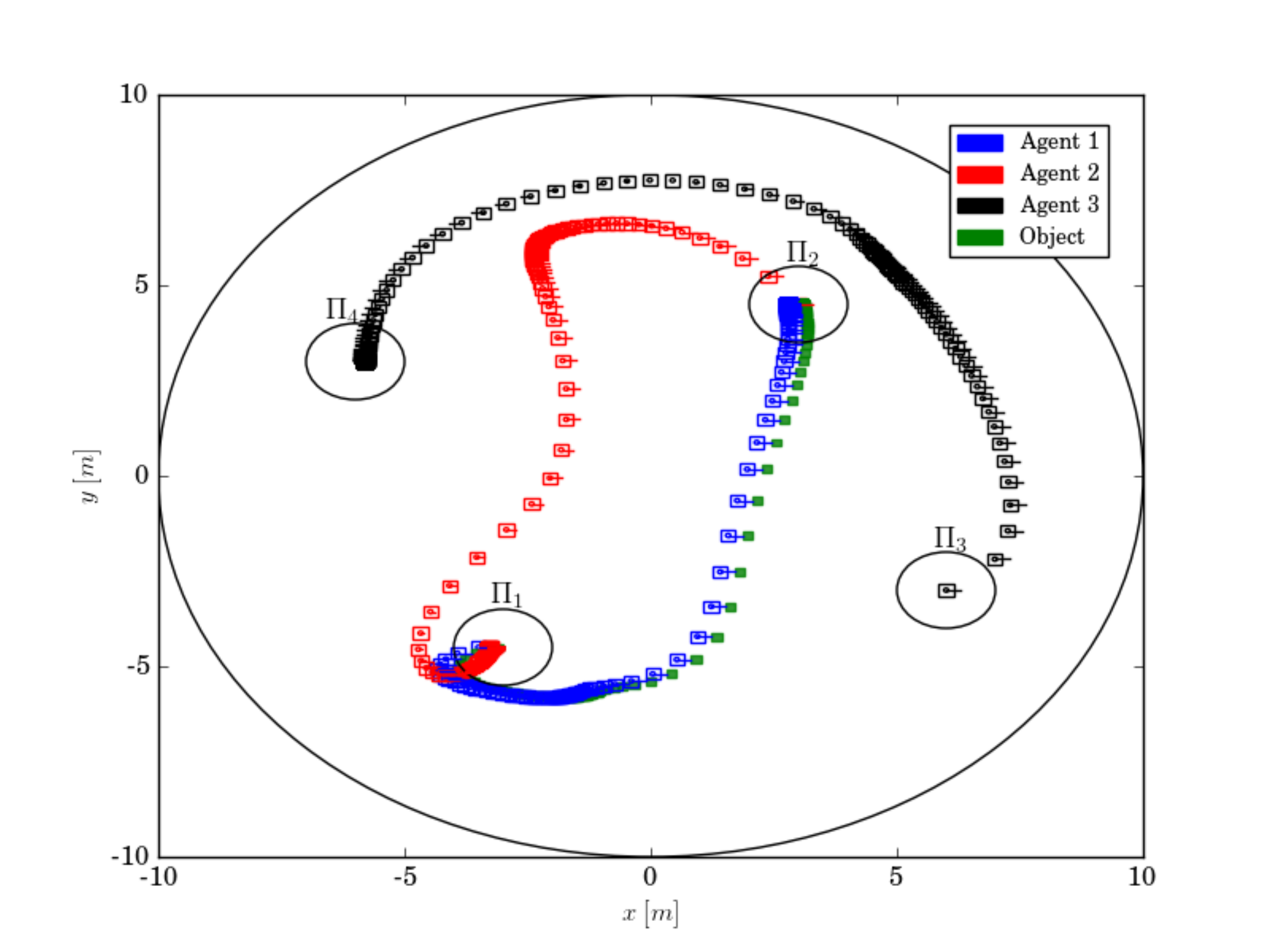}
\caption{The actions $\pi_1 \xrightarrow{T}_{1,1}\pi_2$, $\pi_2 \rightarrow_2\pi_1$ and $\pi_3 \rightarrow_3\pi_4$.  \label{fig:3 agents}}
\end{figure}
\usetikzlibrary{arrows}
\usetikzlibrary{shapes}
\definecolor {darkgreen}{rgb}{0,0.5,0}
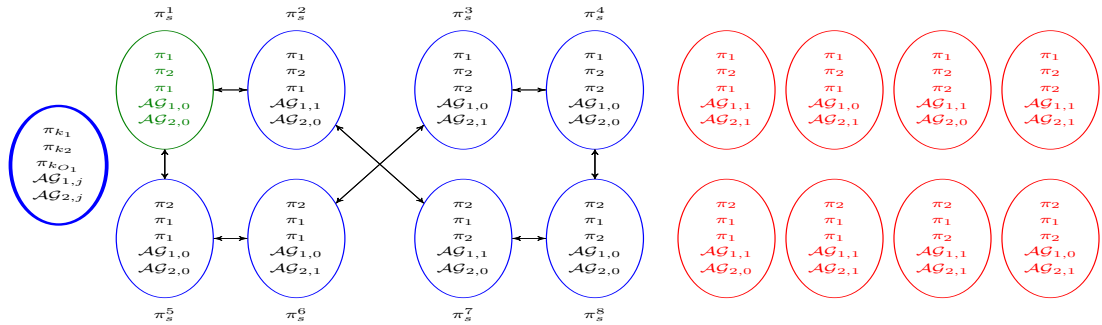
\begin{figure*}
\centering
\resizebox {0.8\textwidth}{0.18\textheight} {
\begin{tikzpicture}[->,>=stealth']

\tikzstyle{every node}=[draw=blue, ellipse, 
    align=center]
\node[label= {\small $\pi^1_{s} $}][darkgreen] (a) { \small $\pi_1$\\  \small $\pi_2$\\  \small $\pi_1$ \\ \small $\mathcal{AG}_{1,0}$\\ \small$\mathcal{AG}_{2,0}$ };
\node[ultra thick][left=45pt, below=10pt] (z) at (a) { \small $\pi_{k_1}$\\  \small $\pi_{k_2}$\\  \small $\pi_{k_{\scriptscriptstyle O_1}}$ \\ \small $\mathcal{AG}_{1,j}$\\ \small$\mathcal{AG}_{2,j}$ };
\node[label= {\small $\pi^2_{s} $}][right=35pt] (b) at (a) {\small$\pi_1$\\ \small$\pi_2$\\ \small $\pi_1$ \\ \small $\mathcal{AG}_{1,1}$\\ \small$\mathcal{AG}_{2,0}$ };
\node[label= {\small $\pi^3_{s} $}][right=50pt] (c) at (b) {\small$\pi_1$\\ \small$\pi_2$\\ \small $\pi_2$ \\ \small $\mathcal{AG}_{1,0}$\\ \small$\mathcal{AG}_{2,1}$ };
\node[label= {\small $\pi^4_{s} $}][right=35pt] (d) at (c) {\small$\pi_1$\\ \small$\pi_2$\\ \small $\pi_2$ \\ \small $\mathcal{AG}_{1,0}$\\ \small$\mathcal{AG}_{2,0}$ };
\node[red][right=35pt] (e) at (d) {\small$\pi_1$\\ \small$\pi_2$\\  \small$\pi_1$ \\ \small $\mathcal{AG}_{1,1}$\\ \small$\mathcal{AG}_{2,1}$ };
\node[red][right=25pt] (f) at (e) {\small$\pi_1$\\ \small$\pi_2$\\ \small $\pi_1$ \\ \small $\mathcal{AG}_{1,0}$\\ \small$\mathcal{AG}_{2,1}$ };
\node[red][right=25pt] (g) at (f) {\small$\pi_1$\\ \small$\pi_2$\\ \small $\pi_2$ \\ \small $\mathcal{AG}_{1,1}$\\ \small$\mathcal{AG}_{2,0}$ };
\node[red][right=25pt] (h) at (g) {\small$\pi_1$\\ \small$\pi_2$\\ \small $\pi_2$ \\ \small $\mathcal{AG}_{1,1}$\\ \small$\mathcal{AG}_{2,1}$ };

\node[label= below:{\small $\pi^5_{s} $}][below=60pt] (i) at (a) {\small$\pi_2$\\ \small$\pi_1$\\ \small $\pi_1$ \\ \small $\mathcal{AG}_{1,0}$\\ \small$\mathcal{AG}_{2,0}$ };
\node[label= below:{\small $\pi^6_{s} $}][right=35pt] (j) at (i) {\small$\pi_2$\\ \small$\pi_1$\\ \small $\pi_1$ \\ \small $\mathcal{AG}_{1,0}$\\ \small$\mathcal{AG}_{2,1}$ };
\node[label= below:{\small $\pi^7_{s} $}][right=50pt] (k) at (j) {\small$\pi_2$\\ \small$\pi_1$\\ \small $\pi_2$ \\ \small $\mathcal{AG}_{1,1}$\\ \small$\mathcal{AG}_{2,0}$ };
\node[label= below:{\small $\pi^8_{s} $}][right=35pt] (l) at (k) {\small$\pi_2$\\ \small$\pi_1$\\ \small $\pi_2$ \\ \small $\mathcal{AG}_{1,0}$\\ \small$\mathcal{AG}_{2,0}$ };
\node[red][right=35pt] (m) at (l) {\small$\pi_2$\\ \small$\pi_1$\\ \small $\pi_1$ \\ \small $\mathcal{AG}_{1,1}$\\ \small$\mathcal{AG}_{2,0}$ };
\node[red][right=25pt] (n) at (m) {\small$\pi_2$\\ \small$\pi_1$\\ \small $\pi_1$ \\ \small $\mathcal{AG}_{1,1}$\\ \small$\mathcal{AG}_{2,1}$ };
\node[red][right=25pt] (o) at (n) {\small$\pi_2$\\ \small$\pi_1$\\ \small $\pi_2$ \\ \small $\mathcal{AG}_{1,1}$\\ \small$\mathcal{AG}_{2,1}$ };
\node[red][right=25pt] (p) at (o) {\small$\pi_2$\\ \small$\pi_1$\\ \small $\pi_2$ \\ \small $\mathcal{AG}_{1,0}$\\ \small$\mathcal{AG}_{2,1}$ };

\path[every node/.style={font=\sffamily\small}]
    (a) edge node [right] {} (b)
    (b) edge node [right] {} (a)

    (a) edge node [right] {} (i)
    (i) edge node [right] {} (a)

    (i) edge node [right] {} (j)
    (j) edge node [right] {} (i)

    (b) edge node [right] {} (k)
    (k) edge node [right] {} (b)
    
    (l) edge node [right] {} (k)
    (k) edge node [right] {} (l)
    
	(c) edge node [right] {} (d)
	(d) edge node [right] {} (c)
	
	(d) edge node [right] {} (l)
	(l) edge node [right] {} (d)
	
	(c) edge node [right] {} (j)
	(j) edge node [right] {} (c);
\end{tikzpicture}
}
\caption{The transition system $\mathcal{TS}$. The information in each state is depicted according to the state with thick blue color. The initial state is colored with green whereas the red states are not reachable. \label{fig:simulation_TS}}
\end{figure*}

\begin{figure*}
\centering
\subfloat[]{\includegraphics[width = 0.45\textwidth,height = 0.2\textheight]{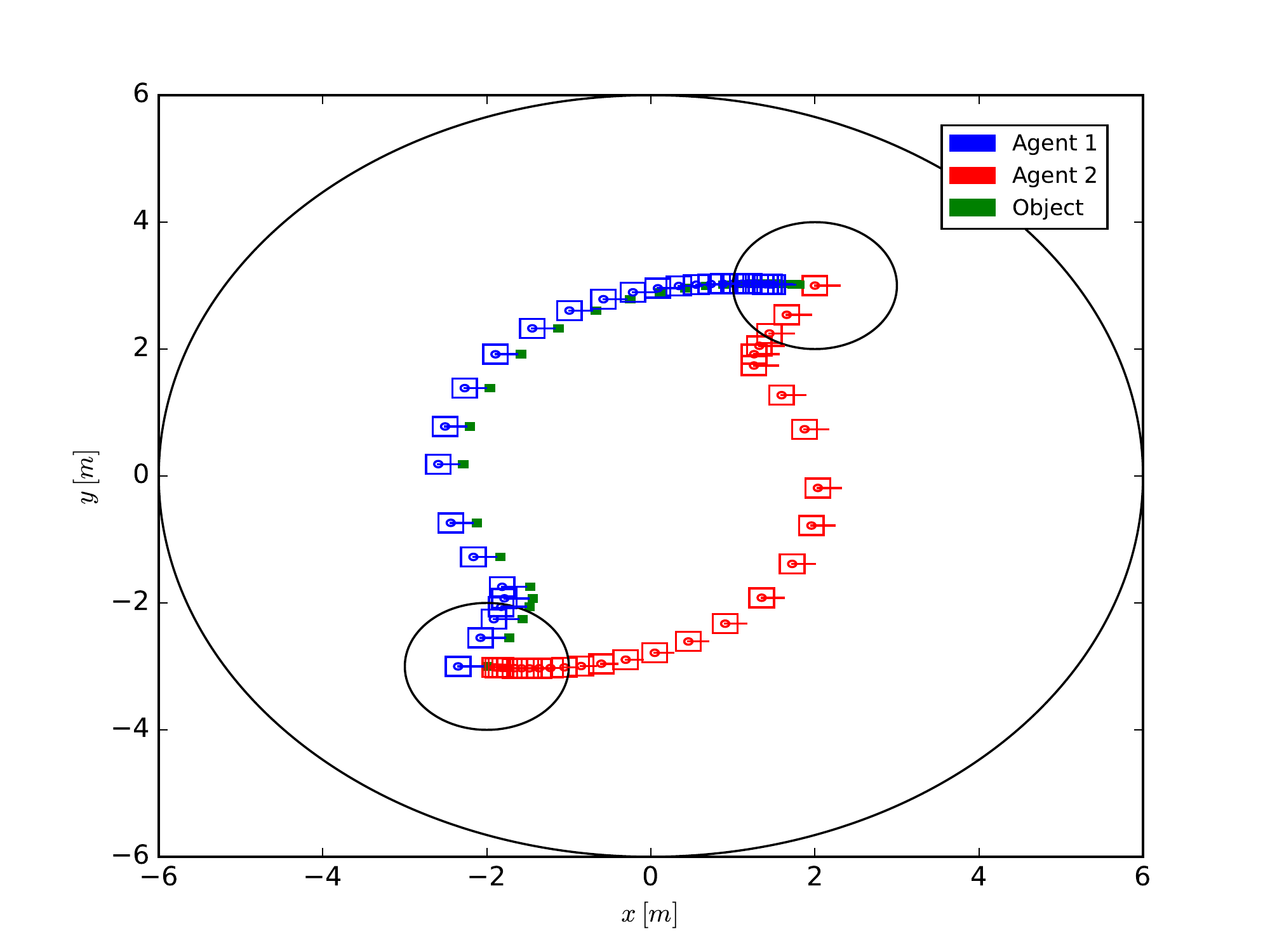}}
\quad
\subfloat[]{\includegraphics[width = 0.45\textwidth,height = 0.2\textheight]{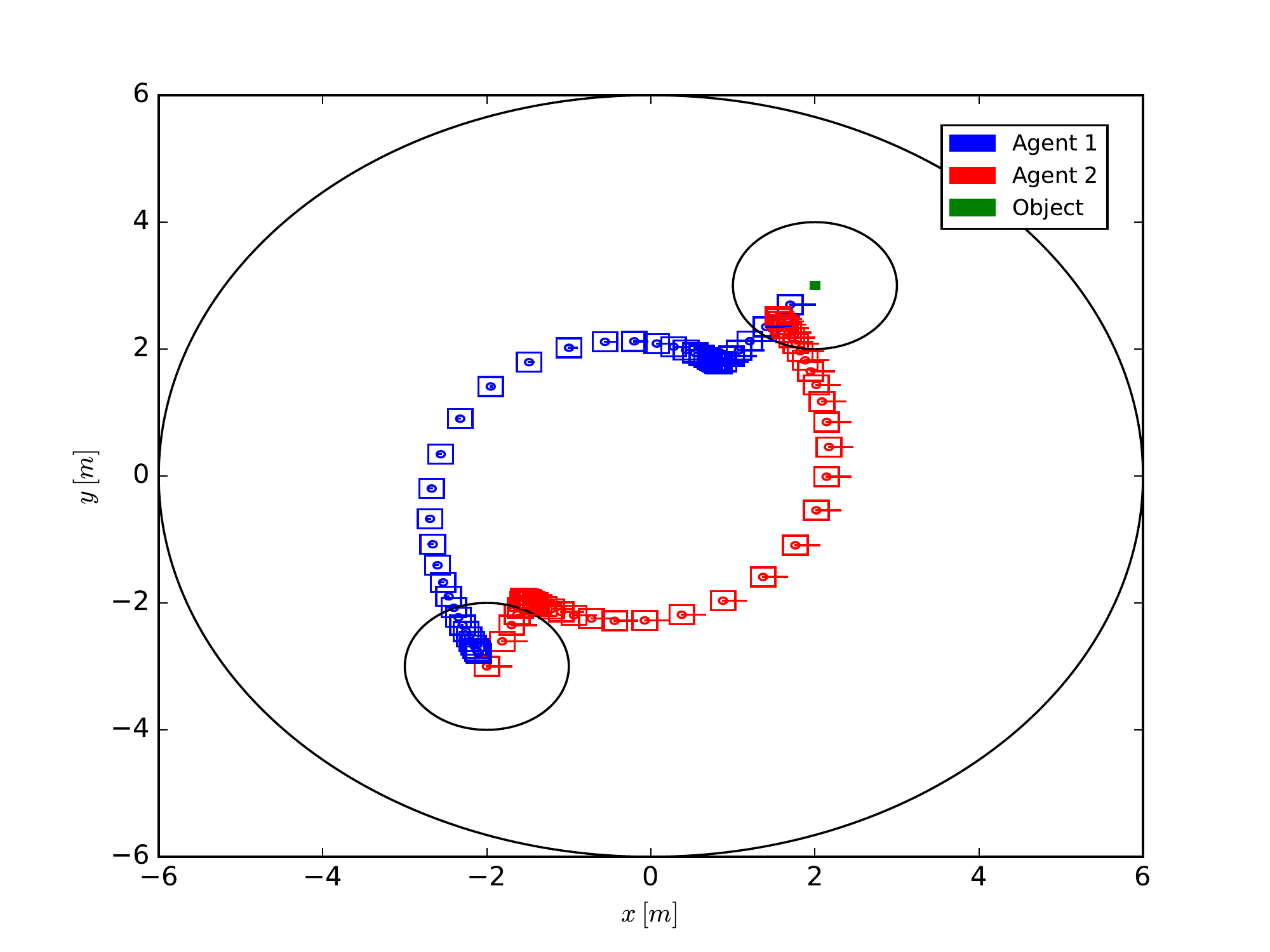}}
\quad
\caption{The transitions $\pi_s^2\to_s\pi_s^7$ and $\pi_s^8\to_s\pi_s^4$ in (a) and (b), respectively.  \label{fig:sim_transitions}}
 \end{figure*}
 
\section{Simulation Results}
Firstly, to demonstrate the efficiency of the continuous control laws \eqref{eq:control_protocol}, we consider a workspace with $\boldsymbol{p}_0 = [0,0,0]^T\text{m}, r_0 = 10\text{m}$, with $K=4$ regions, $\boldsymbol{p}_{\pi_1} = [-4,-4.5,0.2]^T\text{m}$, $\boldsymbol{p}_{\pi_2} = [4,4.5,0.2]^T\text{m}$, $\boldsymbol{p}_{\pi_3} = [6,-4,0.2]^T\text{m}$, $\boldsymbol{p}_{\pi_4} = [-6,4,0.2]^T\text{m}$, $r_{\pi_1}=r_{\pi_2}=r_{\pi_3}=r_{\pi_4}=1\text{m}$, $M=1$ object and $N=3$ of agents. The object is a rigid cube of dimensions $0.1\times0.1\times0.1 \ \text{m}^3$ and each agent consists of a cubic mobile base of dimensions $0.3\times0.3\times0.3 \ \text{m}^3$, able to move on the $x-y$ plane, and two rigid rectangular links of dimensions $0.05\times0.05\times0.3 \ \text{m}^3$ connected by a cylindrical joint rotating around the negative $y$-axis. The generalized variables for each agent are taken as $\boldsymbol{q}_i = [x_{c_i}, y_{c_i}, \theta_i]^T\in\mathbb{R}^3, i\in\{1,2,3\}$, where $[x_{c_i}, y_{c_i}]^T$ is the base's center of mass and $\theta_i$ is the joint's angle. The object's and each agent's volume are approximated by a total of $1$ and $6$ spheres, respectively. The initial conditions are taken such that $\mathcal{A}_1(\boldsymbol{q}_1(0)), \mathcal{O}_{1}(\boldsymbol{x}_{\scriptscriptstyle O_1}(0))\in\pi_1,\mathcal{A}_2(\boldsymbol{q}_2(0))\in\pi_2, \mathcal{A}_3(\boldsymbol{q}_3(0))\in\pi_3$ and $\mathcal{AG}_{1,1}(0) = \top$. We perform a simulation of the actions $\pi_1 \xrightarrow{T}_{1,1}\pi_2$, $\pi_2 \rightarrow_2\pi_1$ and $\pi_3 \rightarrow_3\pi_4$, using \eqref{eq:control_protocol}, with $\mathcal{Z} = \{2,3\}, \mathcal{V}=\{1\}, \kappa = 7, k_1 = k_2 = k_3 = 30, d_1=d_2=d_3=10$. The result is depicted in Fig. \ref{fig:3 agents}, where the agents perform successfully their actions, without colliding with each other.

Next, to demonstrate the overall hybrid control protocol, we consider a simplified scenario involving $N = 2$ of the aforementioned agents, $M=1$ object in a workspace with $\boldsymbol{p}_0 = [0,0,0]^T\text{m}, r_0 = 6\text{m}$, and $K=2$ regions of interest $\pi_1,\pi_2$, with $\boldsymbol{p}_{\pi_1} = [-2,-3,0.2]^T\text{m}$, $\boldsymbol{p}_{\pi_2} = [2,3,0.2]^T\text{m}, r_{\pi 1}=r_{\pi 2}=1\text{m}$. The initial conditions are taken such that $\mathcal{A}_1(\boldsymbol{q}_1(0)),\mathcal{O}_1(\boldsymbol{x}_{\scriptscriptstyle O_1}(0))\in\pi_1,\mathcal{A}_2(\boldsymbol{q}_2(0))\in\pi_2$ and $\Omega^{1,*}(\boldsymbol{q}_1(0))\cap\Omega^{1,*}_{\scriptscriptstyle O}(\boldsymbol{x}_{\scriptscriptstyle O_1}(0)) = \emptyset$. 

The resulting $\mathcal{TS}$ is pictured in Fig. \ref{fig:simulation_TS}, where we show each state $\pi^m_{s}$ in the form $(\pi_{k^m_1},\pi_{k^m_2},\pi_{k^m_{\scriptscriptstyle O_1}},\mathcal{AG}_{1,j},\mathcal{AG}_{2,j})$, as depicted with thick blue color in the figure, with $j\in\{0,1\}$. The initial state is
$\Pi_s^{\text{init}} = (\pi_{k^1_1}$,$\pi_{k^1_2}$,$\pi_{k^1_{\scriptscriptstyle O_1}}$,$w^1_1$,$w^1_2) = (\pi_1$,$\pi_2$,$\pi_1$,$\mathcal{AG}_{1,0}$,$\mathcal{AG}_{2,0})$ (depicted with green color in the figure). Note that, due to our restriction that no more than one agent is allowed to in the same region, the number of states is reduced from $K^{N+M}(M+1)^N = 32$ to $16$. Moreover, since an agent $i$ cannot have a grasp with object $j$ if $\pi_{k^m_i}\neq \pi_{k^m_{\scriptscriptstyle O_j}}, \forall m\in\mathbb{N}$, some states are not reachable (depicted with red in the figure), and thus the number of states is further reduced to $8$. We also consider the atomic propositions $\Psi_1 = \{\text{``red"}, \text{``blue"}\}, \Psi_2 = \{\text{``green"}, \text{``yellow"}\}$ and $\Psi_{\scriptscriptstyle O_1} = \{\text{``Goal}_1\text{"}, \text{`
`Goal}_2\text{"}\}$, with $\mathcal{L}_1(\pi_1) = \{\text{``red"}\},\mathcal{L}_1(\pi_2) = \{\text{``blue"}\},\mathcal{L}_2(\pi_1) = \{\text{``green"}\},\mathcal{L}_2(\pi_2) = \{\text{``yellow"}\}$, and $\mathcal{L}_{\scriptscriptstyle O_1}(\pi_1) = \{\text{``Goal}_1\text{"}\},\mathcal{L}_{\scriptscriptstyle O_1}(\pi_2) = \{\text{``Goal}_2\text{"}\}$. The formulas to be satisfied by the agents and the object are the following: $\phi_1 = \square\lozenge(\text{``red"}\land\lozenge\text{``blue"}),\phi_2 = \square\lozenge(\text{``green"}\land\lozenge\text{``yellow"})$ and $\phi_{\scriptscriptstyle O_1} = \square\lozenge(\text{``Goal}_1\text{"}\land\lozenge\text{``Goal}_2\text{"})$. By following the procedure described in Section \ref{sec:high level plan}, we obtain a path satisfying $\phi = \phi_1\land\phi_2\land\phi_{\scriptscriptstyle O_1}$ as $\pi^1_s\pi_s^2(\pi_s^7\pi_s^8\pi_s^4\pi_s^3\pi_s^6\pi_s^5\pi_s^1\pi_s^2)^\omega$, which includes transitions, grasping/releasing as well as transportation actions from both agents. Fig. \ref{fig:sim_transitions} depicts two indicative transitions, namely, $\pi_s^2\to_s\pi_s^7$ and $\pi_s^8\to_s\pi_s^4$.

\section{Conclusion}
We have presented a novel hybrid control framework for the motion planning of a system comprising of $N$ agents and $M$ objects. Future works will address decentralization of the framework as well as cooperative transportation of the objects by agents with limited sensing information.
\bibliography{ifacconf}             
                                                   







\end{document}